\documentclass[12pt]{spieman}  
\usepackage{amsmath,amsfonts,amssymb}
\usepackage{graphicx}
\usepackage{setspace}
\usepackage{tocloft}
\usepackage{multirow}
\usepackage[final]{pdfpages}
\usepackage{booktabs} 
\usepackage{siunitx} 
\usepackage{subcaption}
\usepackage{textcomp}

\usepackage{lineno}

\usepackage{hyperref}
\hypersetup{
    colorlinks=true,
    linkcolor=blue,
    filecolor=magenta,      
    urlcolor=cyan,
}

\title{WALOP-South: A Four Camera One Shot Imaging Polarimeter for PASIPHAE Survey. Paper I - Optical Design}

\author[a*]{Siddharth Maharana}
\author[b,c]{John A. Kypriotakis}
\author[a,b,e]{A. N. Ramaprakash}
\author[a]{Chaitanya Rajarshi}
\author[a]{Ramya M. Anche}
\author[h]{Shrish}
\author[b,c,i]{Dmitry Blinov}
\author[g]{Hans Kristian Eriksen}
\author[h]{Tuhin Ghosh}
\author[g]{Eirik Gjerløw}
\author[b,c]{Nikolaos Mandarakas}
\author[f]{Georgia V. Panopoulou}
\author[b,c]{Vasiliki Pavlidou}
\author[e]{Timothy J. Pearson}
\author[b,c]{Vincent Pelgrims}
\author[d,j]{Stephen B. Potter}
\author[e]{Anthony C. S. Readhead}
\author[b,c]{Raphael Skalidis}
\author[b,c]{Konstantinos Tassis}
\author[g]{Ingunn K. Wehus}

\affil[a]{Inter-University Centre for Astronomy and Astrophysics, Post bag 4, Ganeshkhind, Pune, 411007, India}
\affil[b]{Institute of Astrophysics, Foundation for Research and Technology-Hellas, Voutes, 70013 Heraklion, Greece}

\affil[c]{Department of Physics, University of Crete, Voutes, 70013 Heraklion, Greece}

\affil[d]{South African Astronomical Observatory, PO Box 9, Observatory, 7935, Cape Town, South Africa}

\affil[e]{Cahill Center for Astronomy and Astrophysics, California Institute of Technology, Pasadena, CA, 91125, USA}

\affil[f]{Hubble Fellow, California Institute of Technology, Pasadena, CA 91125, USA}

\affil[g]{Institute of Theoretical Astrophysics, University of Oslo, P.O. Box 1029 Blindern, NO-0315 Oslo, Norway}

\affil[h]{School of Physical Sciences, National Institute of Science Education and Research, HBNI, Jatni 752050, Odisha, India}

\affil[i]{Astronomical Institute, St. Petersburg State University, 198504, St. Petersburg, Russia}

\affil[j]{Department of Physics, University of Johannesburg, PO Box 524, Auckland Park 2006, South Africa}

\cftpagenumbersoff{figure}
\cftpagenumbersoff{table} 
\begin{document} 
\maketitle

\begin{abstract}
The WALOP-South instrument will be mounted on the 1~m SAAO telescope in South Africa as part of the PASIPHAE program to carry out a linear imaging polarization survey of the Galactic polar regions in the optical band. Designed to achieve polarimetric sensitivity of 0.05\% across a $35\times 35$ ~arcminute field of view, it will be capable of measuring the Stokes parameters I, $q$ and $u$ in a single exposure in the SDSS-r broadband and narrowband filters between $0.5~{\mu}m - 0.7~{\mu}m$.  For each measurement, four images of the full field corresponding to linear polarization angles of $0^{\circ}$, $45^{\circ}$, $90^{\circ}$ and $135^{\circ}$ in the instrument coordinate system will be created on four detectors from which the Stokes parameters can be found using differential photometry. In designing the optical system, major challenges included correcting for the dispersion introduced by large split angle Wollaston Prisms used as analysers as well as other aberrations from the entire field to obtain imaging quality PSF at the detector. We present the optical design of the WALOP-South instrument which overcomes these challenges and delivers near seeing limited PSFs for the entire field of view.
\end{abstract}


{\noindent \footnotesize\textbf{*}Siddharth Maharana,  \linkable{sidh@iucaa.in} }


\section{Introduction}

Optical polarimetry in nighttime astronomy began in the 1940's with Hiltner\cite{Hiltner165} and Hall's\cite{Hall166} pioneering work on polarization measurement of stars, leading to the serendipitous discovery of interstellar polarization. Since then, optical polarimetry has been used as an essential tool by astronomers to make progress in understanding various classes of objects such as stars, active galactic nuclei, exoplanets and protoplanetary disks\cite{10.1111/j.1468-4004.2006.47331.x, Scarrott-1991}. With advancements in detectors, polarization optics hardware as well as associated control systems, optical polarimeters with polarimetric sensitivity\footnote{We define polarimetric sensitivity ($s$) as the least value and change of linear polarization which the instrument can measure, without correction for the cross-talk and instrumental polarization of the instrument. $s$ is a measure of the internal noise and random systematics of the instrument due to the optics. Polarimetric accuracy ($a$) is the measure of closeness of the predicted polarization of a source to the real value after applying the above corrections using calibration techniques (Section~\ref{calibration}).} ($s$) better than $10^{-5}$ in $p$ (fractional polarization) have been made, eg. HIPPI-2 \cite{Bailey_2020} and DIPOL-2\cite{DIPOL2}. But most polarimeters built to date have limited field of view (FOV) of about $1\times1$~arcminute or less. So while we have extensive  polarization measurements of many of the above mentioned individual classes of objects, a large area ($>1000$~square degrees) continuous optical polarization map of the sky has been unavailable. Existing large stellar polarization catalogues include measurements of nearly 3600 and 10000 individual stars by Berdyugin et al.\cite{Berdyugin_catalog} and Heiles\cite{Heiles_2000} respectively in the optical wavelengths, and a continuous map of $76$~square degrees of the galactic plane by the GPIPS survey\cite{GPIPS_Survey} in the near infrared wavelengths. 

\par Using two WALOP (Wide-Area Linear Optical Polarimeter) polarimeters as survey instruments, the \href{http://pasiphae.science/}{PASIPHAE} (Polar-Areas Stellar Imaging Polarization High Accuracy Experiment) program aims to create a unique optopolarimetric map of the sky. Such a map will enable astronomers to pursue answers to many open questions related to the physics of dust and magnetic fields in the interstellar medium (ISM), which together are the main source of starlight polarization. A detailed description of the motivation and scientific objectives of the PASIPHAE survey can be found in the PASIPHAE white paper by Tassis et al.\cite{tassis2018pasiphae}. Here we mention key highlights of the program:
\begin{itemize}
    \item[--] In the northern and southern Galactic polar regions, cover $>4000$~square degrees of the sky and measure polarization of about $10^{6}$ stars with polarimetric accuracy (a) of 0.1~\%. Current optical polarization catalogues have polarization measurements of around $10^{4}$ stars\cite{Heiles_2000}.
    \item[--] The survey will be simultaneously carried out from the \href{https://www.saao.ac.za/}{1~m telescope} at SAAO's (South African Astronomical Observatory) Sutherland Observatory, South Africa in the southern hemisphere and the \href{https://skinakas.physics.uoc.gr/}{1.3~m telescope} at Skinakas Observatory, Greece in the northern hemisphere by using the WALOP-South and WALOP-North instruments, respectively.
    \item[--] Using stellar polarimetry in conjugation with the \href{https://sci.esa.int/web/gaia}{GAIA} mission's distance measurements of stars, carry out magnetic field and dust cloud tomography of the ISM. The methodology to create such a map has been discussed by Panopoulou et al.\cite{Panopoulou_2019}.
    \item[--] The main science goal of the PASIPHAE program is to aid CMB B-mode polarization searches by identifying clean patches of sky suitable for the search as well as to improve the foreground emission models by combining PASIPHAE's tomographic map with polarized dust emission data. 
    \item[--] Some of the secondary science goals are- (a) improving understanding of interstellar dust by testing various physical models of grain alignment and size distribution, (b) finding the paths of ultra high-energy cosmic rays though the Galaxy, (c) creating a catalogue of intrinsically polarized stars and finding associated systematic correlations with properties like the spectral type and stage of the stars.
\end{itemize}

\par Of the two WALOP instruments, WALOP-South is scheduled to be commissioned first in \textcolor{blue}{2021}. Both the WALOPs are currently under development at IUCAA, Pune. The unique scientific goals of the PASIPHAE survey lead to a set of very challenging design and performance requirements for the optical system of WALOP instruments.

\par This paper is the first of a series of papers describing the design, polarization modelling, calibration and on-sky performance of the WALOP-South instrument\cite{walop_s_spie_2020}. In this paper, we present the complete optical design of the instrument. The optical design of WALOP-North is similar to that of WALOP-South- the differences are due to the differences in the telescope optics (they have different f-numbers) as well as opto-mechanical interfaces.
Section~\ref{techgoals} describes the technical goals of the instrument as driven by the scientific objectives of the PASIPHAE survey and the challenges in realizing them. Section~\ref{design} explains the overall optical design and Section~\ref{pol_design} gives a detailed description of the architecture and working of the polarization analyzer subsystem of the instrument, referred to as the polarizer assembly in this paper, which is the most complex and novel subsystem of the instrument's optical system. In Section~\ref{performance}, we show the performance of the design and Section~\ref{conclusion} contains our conclusions and observations regarding the optical design of WALOP-South and its possible application to the design of other wide field polarimeters. 
In addition to the WALOP-South optical system, we designed new baffles for the telescope's mirrors to accommodate the wide FOV and an auto-guider camera. These are presented in Appendix~\ref{baffles} and \ref{guider} respectively.
\section{Technical Requirements of WALOP-South instrument}\label{techgoals}
\par Based on PASIPHAE survey goals as well as the current state of the art optical polarimeter design technology and understanding, optical design goals for WALOP-South instrument were decided. These goals are captured in Table~\ref{techtable}. The design goals for WALOP-North are same as WALOP-South.
\par Most stars at high galactic latitude are expected to have $p$~ $<0.5\%$ due to lower dust content\cite{Skalidis} in these regions. To accomplish the scientific objectives of the PASIPHAE program, a wide field polarimeter with high accuracy ($a$) and high sensitivity ($s$) is needed. As the survey aims to measure $p$ with $a =< 0.1~\%$ as limited by photon noise, we aim to limit $s =< 0.05~\%$.

\par
\begin{figure}
    \centering
    \frame{\includegraphics[scale =0.3]{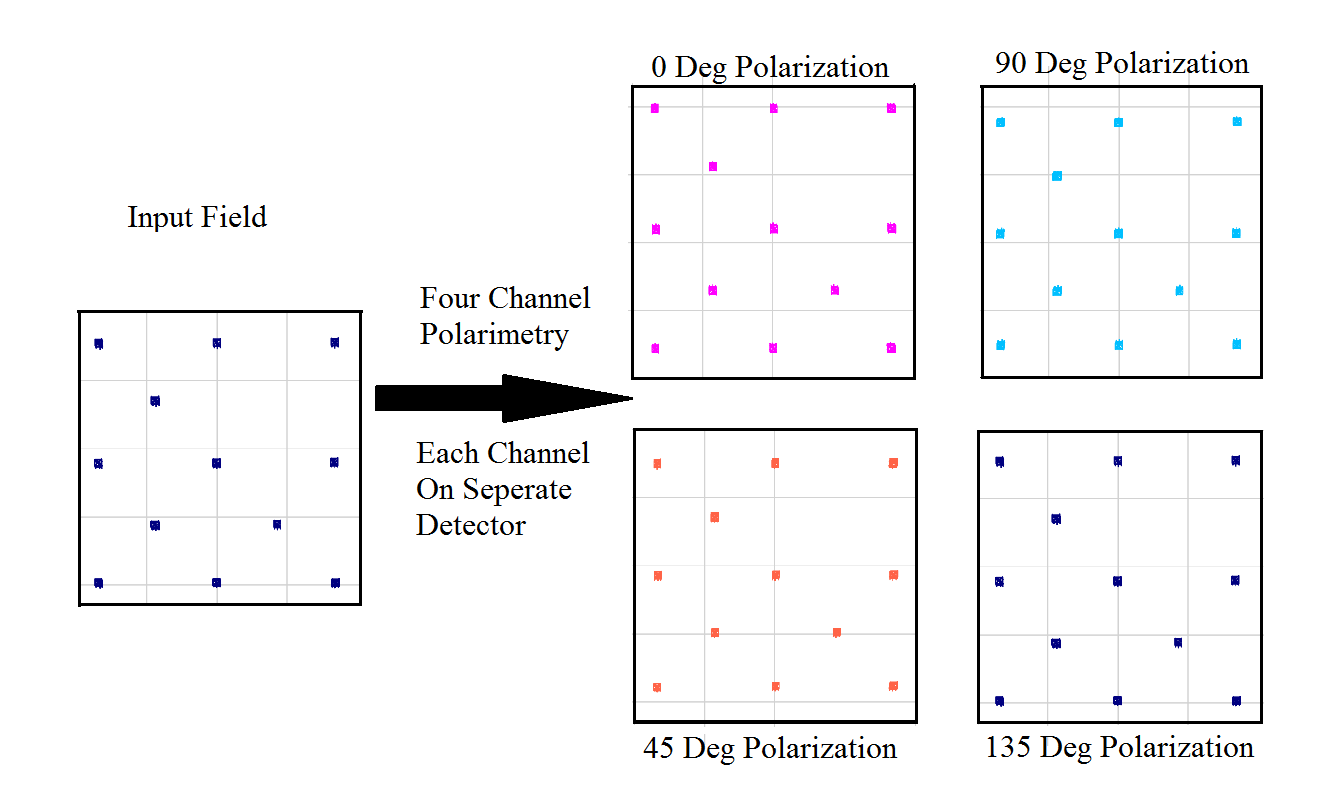}}
    \caption{Concept of four channel imaging with separate camera for each channel. The input field is split into four channels along the $0^{\circ}$, $45^{\circ}$, $90^{\circ}$ and $135^{\circ}$ polarization angles and imaged on four separate detectors without any overlap.}
    \label{four_channel}
\end{figure}

 The polarimeter will be capable of carrying out four channel one-shot linear imaging polarimetry in R broadband filter and narrowband filters lying between $0.5~{\mu}m$ to $0.7~{\mu}m$. Previous optical polarimeters like  RoboPol\cite{robopol} have demonstrated the benefit of one-shot linear polarimetry, which avoids instrumental noise due to rotating components like Half Wave Plates (HWP) as well as changing sky conditions such as airmass during the exposures to obtain better polarimetric sensitivity. Each of the four channels is imaged on a separate detector or detector area. A schematic of this concept is shown in Fig~\ref{four_channel}, where after passing through the polarimeter, four images of the input field along the $0^{\circ}$, $45^{\circ}$, $90^{\circ}$ and $135^{\circ}$ polarization angles are imaged on separate detectors. This approach has three major advantages over conventional polarimeters which image all two or four channels on the same detector area. Firstly, the background sky is an extended object, so imaging the four channels on different detectors reduces the sky background by a factor of four by avoiding overlap of ordinary and extraordinary images from adjacent sky regions. Second, there is no intermixing of images from different channels, enabling more accurate photometry and polarimetry in sky regions with higher stellar density. Finally, this approach allows extended object imaging polarimetry for a large field, which has not been possible without using slit masks in previous polarimeters like RoboPol\cite{robopol} and IMPOL\cite{impol}, leading to obscuration of large regions of the FOV.

\par The initial PASIPHAE survey goal with WALOP-South is to cover 1000 square degrees, mainly covering the areas targeted by upcoming CMB B-mode search missions. 
With 200 available nights on the telescope per year, an average of 8 hours per night and 70\% observation efficiency, we can cover a 1000 square degrees region in 16 months with a FOV of $30\times30$~arcminutes and measure $p$ with $a =< 0.1~\%$ for R $=< 14.0$ mag stars (with the obtained FOV of $35\times35$~arcminutes, the estimated survey period becomes 14 months).

\par The median seeing FWHM (full width half maximum) at the Sutherland Observatory is 1.5~arcseconds. Imaging the FOV on a $4k\times4k$ detector with pixel size of $15~{\mu}m$ gives a plate scale of 0.45~arcseconds per pixel, allowing better than Nyquist sampling. The PSF (point spread function) should be close to seeing limited for median seeing conditions. While a larger PSF helps in spreading photons over more pixels and washing out pixel to pixel sensitivity variation, it comes at the cost of a higher measurement uncertainty from increased sky background.

\begin{table}
    \centering
    \begin{tabular}{|c|c|c|}
        \hline
        \textbf{Sl. No}. & \textbf{Parameter} & \textbf{Technical Goal} \\
         \hline
        1 & Polarimetric Sensitivity ($s$) & 0.05~\%\\
         \hline
        2 & Polarimeter Type & Four Channel One-Shot Linear Polarimetry \\
         \hline
        3 & Number of Cameras & 4 (One Camera for Each Arm)\\
         \hline
        4 & Field of View & $30\times30$~arcminutes\\
         \hline
        5 & Detector Size & $4k\times4k$ (Pixel Size = $15~{\mu}m$) \\
        \hline
        6 & No. of Detectors & 4 \\
        \hline
        7 & Primary Filter & SDSS-r \\
        \hline
        8 & Imaging Performance & Close to seeing limited PSF \\
         \hline
        9 & Stray and Ghost Light Level & Brightness less than sky brightness per pixel.\\
        \hline
    \end{tabular}
    \caption{Design goals for WALOP-South instrument optical system.}
    \label{techtable}
\end{table}

\par The stray light from objects outside the FOV and ghost light from field objects due to reflections from telescope baffles/instrument optics should be reduced to a level where their intensities at the detector will introduce less than 0.05\% polarization (less than $s$). While this cannot be realised for very bright stars and moon if they are near the FOV, for most sources this can be achieved by keeping their stray/ghost light levels below the sky background.

\subsection{Challenges in making the WALOP-South Optical Design}
The main challenge in building WALOP-South is obtaining $s =<0.05\%$ across the FOV. Wollaston Prisms (WP) are the most widely used linear polarization analyzers in optical astronomy due to their high extinction ratios ($>10^{5}$) as well as near symmetrical angular splitting of orthogonal polarization states, making these suitable to be placed at a pupil plane. It was decided to use a double WP system to create a four channel analyzer system (refer to Section~\ref{tradeoff} for a trade-off study between different candidate analyzer systems). While double WPs with separate imaging of four channels on different detector areas have been designed and implemented in astronomical polarimeters in the past, eg. in HOWPol\cite{HOWPol} using WeDoWo prisms\cite{dowedo} and in RoboPol\cite{robopol}, these instruments have smaller FOVs for which WPs with split angles of around $1^{\circ}$ are sufficient. However, for WALOP-South's field size, large split angles of the order $5^{\circ}$ to $10^{\circ}$ are required. WPs with such large split angles introduce large spectral dispersion in broadband filters in the split beams due to the dependence of split angle on wavelength.
This is in addition to the usual problems of large aberrations of off-axis objects due to the very wide field.

\section{Instrument Design}\label{design}
\subsection{Telescope And Site Details}
WALOP-South will be mounted on the direct port of the 1~m telescope at SAAO's Sutherland Observatory. Details of the telescope and site is captured in Table~\ref{telescope_details}. The instrument has been designed to perform optimally for the temperature range and seeing conditions at the site.

\begin{table}[ht!]
    \centering
    \begin{tabular}{|c|c|}
    \hline
    \textbf{Parameter} & \textbf{Value} \\
    \hline
    Telescope Type & Cassegrain Focus and Equatorial Mount \\
    \hline
    Primary Mirror Diameter & 1~m \\
    \hline
    Secondary Mirror Diameter & 0.33~m \\
    \hline
    Nominal Telescope f-Number & 16.0 \\
    \hline
    Altitude & 1800~m\\
    \hline
    Median Seeing FWHM & 1.5" \\
    \hline
    Extreme Site Temperatures & $-10^{\circ}~C$ to $40^{\circ}~C$\\
    \hline
    \end{tabular}
    \caption{Telescope and Site Details.}
    \label{telescope_details}
\end{table}

\subsection{Optical Design Overview}
\begin{figure}
    \centering
    \fbox{\includegraphics[scale=0.45]{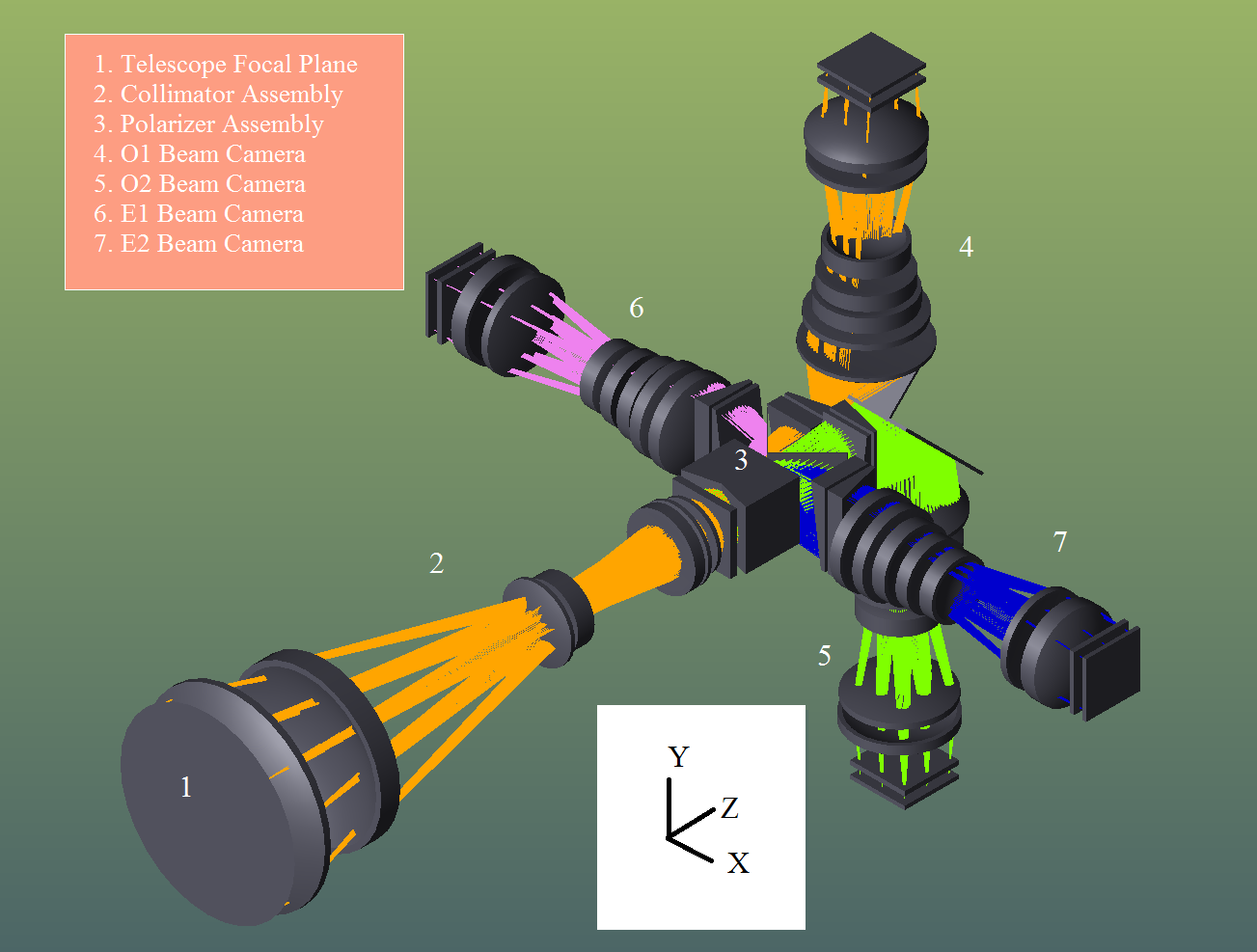}}
    \caption{Complete optical model of the WALOP-South instrument. It accepts the beam for the complete FOV from the telescope focal plane and through the collimator assembly creates a pupil which is then fed to the polarizer assembly. The polarizer assembly splits the pupil beam into four channels and steers them in +/- x and +/- y directions. The O1 and O2 beams correspond to $0^{\circ}$, $45^{\circ}$ polarization while the E1 and E2 beams correspond to $90^{\circ}$ and $135^{\circ}$ polarization. Each channel has its own camera assembly to image the complete field on a $4k\times4k$ detector.}
    \label{WALOP-S}
\end{figure}

The optical model of WALOP-South was created and analyzed using the \href{https://www.zemax.com/}{Zemax}\textsuperscript{\textregistered} optical design software. The complete instrument model is shown in Fig~\ref{WALOP-S}. The instrument can be divided into the following assemblies: a collimator, a polarizer and four cameras (one for each channel). The collimator assembly, beginning from the telescope focal plane is aligned along the z-axis and creates a pupil where the polarizer assembly is placed. The polarizer assembly splits the incoming collimated beam into four channels corresponding to $0^{\circ}$, $45^{\circ}$, $90^{\circ}$ and $135^{\circ}$ polarization angles, which are referred to as O1, O2, E1 and E2 beams respectively. Additionally, this assembly steers the O beams along the +y and -y directions and the E beams along the +x and -x directions. Each channel has its own camera which images the entire field on a $4k\times4k$ detector. The obtained FOV of the instrument is $34.8\times34.8$~arcminutes (see Section~\ref{WP_design} for more details). The key design parameters of the optical design are listed in Table~\ref{op_design_summary}.
\par While creating the optical design, a major consideration was to avoid the use of mirrors and aspheric lenses. Mirrors introduce instrumental polarization due to reflections while aspheric lenses are relatively more difficult to fabricate and align in an optical assembly. In addition to these, the length of the instrument must be restricted to less than $1.5~m$ from the telescope focal plane due to space constraints at the telescope.

\par Almost all the complexity of WALOP-South optical design resides in the architecture and working of the polarizer assembly. Its design and working are explained in detail in Section~\ref{pol_design}. The collimator and camera assemblies are described in Sections~\ref{collimator} and \ref{camera}.

\begin{table}
    \centering
    \begin{tabular}{|c|c|}
        \hline
        \textbf{Parameter} & \textbf{Design Value/Choice}\\
        \hline
        Filter & SDSS-r \\
        \hline
        Telescope F-number & 16.0 \\
        \hline
        Camera F-number & 6.1 \\
        \hline
        Collimator Length & 700~mm\\
        \hline
        Camera Length & 340~mm \\
        \hline
        {Pupil Diameter} & {65~mm} \\
        \hline
        No of lenses in Collimator & 6 \\
        \hline
        No of lenses in Each Camera & 7 \\
        \hline
        Detector Size & $4096\times4096$ \\
        \hline
        Pixel Size & $15~{\mu}m$ \\
        \hline
        Sky Sampling at detector & 0.5"/pixel\\
        \hline
        
    \end{tabular}
    \caption{Values of the key parameters of WALOP-South Optical Design.}
    \label{op_design_summary}
\end{table}

\subsection{Polarizer Assembly}\label{pol_design}

\subsubsection{Design Choices for the Polarizer System}\label{tradeoff}

Our key requirements in deciding on a suitable polarization analyzer design were: (a) to achieve four channel beam-splitting as shown in Figure~\ref{four_channel}, (b) to achieve good PSFs at the detectors from the split beams for the entire FOV,  and (c) to ensure that the split beams have high extinction ratios ($>10^{4}$) to achieve the required $s$. To decide on the most suitable architecture of the polarizer system, we considered four broad types of polarization analyzer systems and a trade-off study was carried out to find the most suitable solution. These were: 

\begin{enumerate}
    \item Twin Wire-Grid Polarization Beam Splitters (PBS).
    \item Twin Glan-Taylor/Thomson Prisms.
    \item Twin Polarization Gratings.
    \item Twin Wollaston Prisms.
\end{enumerate}

\par A single Wire-Grid PBS separates the incident beam into two orthogonal polarization states (parallel and perpendicular to the direction of nano-wires in the PBS) and steers them in orthogonal directions- one polarization is transmitted while the other is reflected. These have been used in two-channel polarimeters like MOPTOP\cite{moptop} to obtain on-sky accuracy of 0.25~\% using a modulating HWP. The advantage of using these is that their beam-splitting performance does not change significantly over large angles of incidence, or over the range of wavelengths of broadband filters like the SDSS-r, which is a major problem with using crystal based polarizers like Wollaston Prisms, as described later. While the transmitted beam can have a very high extinction ratio of $>10^{4}$, the reflected beam has extinction ratio of less than 100, which would compromise the sensitivity of the instrument in a single shot polarimeter like WALOP-South.  
\par The Glan-Taylor Prisms also work in a very similar manner as the Wire-Grid PBS- orthogonal polarization states are split and steered in orthogonal directions. Instead of nano-wires, Glan-Taylor Prisms use anisotropic optical properties of birefringent crystals like calcite to reflect one of the polarizations using total internal reflection. Similar to Wire-Grid PBS, the reflected beam has very low extinction ratio. Additionally, in general, these polarization beam-splitters suffer from severe degradation of polarization performance even for small angles of $1-2^{\circ}$ away from normal incidence, whereas the collimated beam of WALOP-South at pupil has angles of up to $6^{\circ}$. 

\par Polarization gratings (PG)\cite{Packham_2010} have been used in astronomical spectropolarimetry to achieve 0.1~\% accuracy in four channel polarimeters, e.g. WIRC+Pol\cite{Tinyanont2018}. An advantage of PGs for use in large FOV polarimeters is their availability in large aperture sizes, whereas a major problem associated in employing PGs in imaging polarimeters is the large dispersion introduced in the outgoing beams. Also, since PGs separate circular polarizations, a quarter-wave plate (QWP) is needed in front of the PGs to make them separate linear polarization states, making them more suitable for circular polarimetry than linear polarimetry. 

Finally, Wollaston Prism (WP) systems have been used in astronomy\cite{robopol,HOWPol,dowedo,salt_commisioning} as the go-to polarization analyzer system. These separate the orthogonal polarization states (called E and O beams) with high extinction ratios. For WALOP-South's FOV, WPs with large split angles ($>~5^{\circ}$) and apertures are needed to image the four split beams on different detectors. Also, the split angle of a WP depends on the wavelength, and for a broadband filter like the SDSS-r, there is large dispersion in the outgoing beams (Section~\ref{dispersion}, Figure~\ref{WP_dispersion}).

Since our main goal is to carry out sensitive polarimetry, we decided to go with twin Wollaston Prisms as the polarization analyzer system, with each Wollaston Prism having its own HWP in front, similar to the architecture of RoboPol's polarization analyzer system\cite{robopol}. The dispersion introduced by the Wollaston Prism system is corrected downstream in the optical design.

\color{black}

\subsubsection{Polarizer Assembly Overview}
\begin{figure}
    \centering
    \frame{\includegraphics[scale = 0.25]{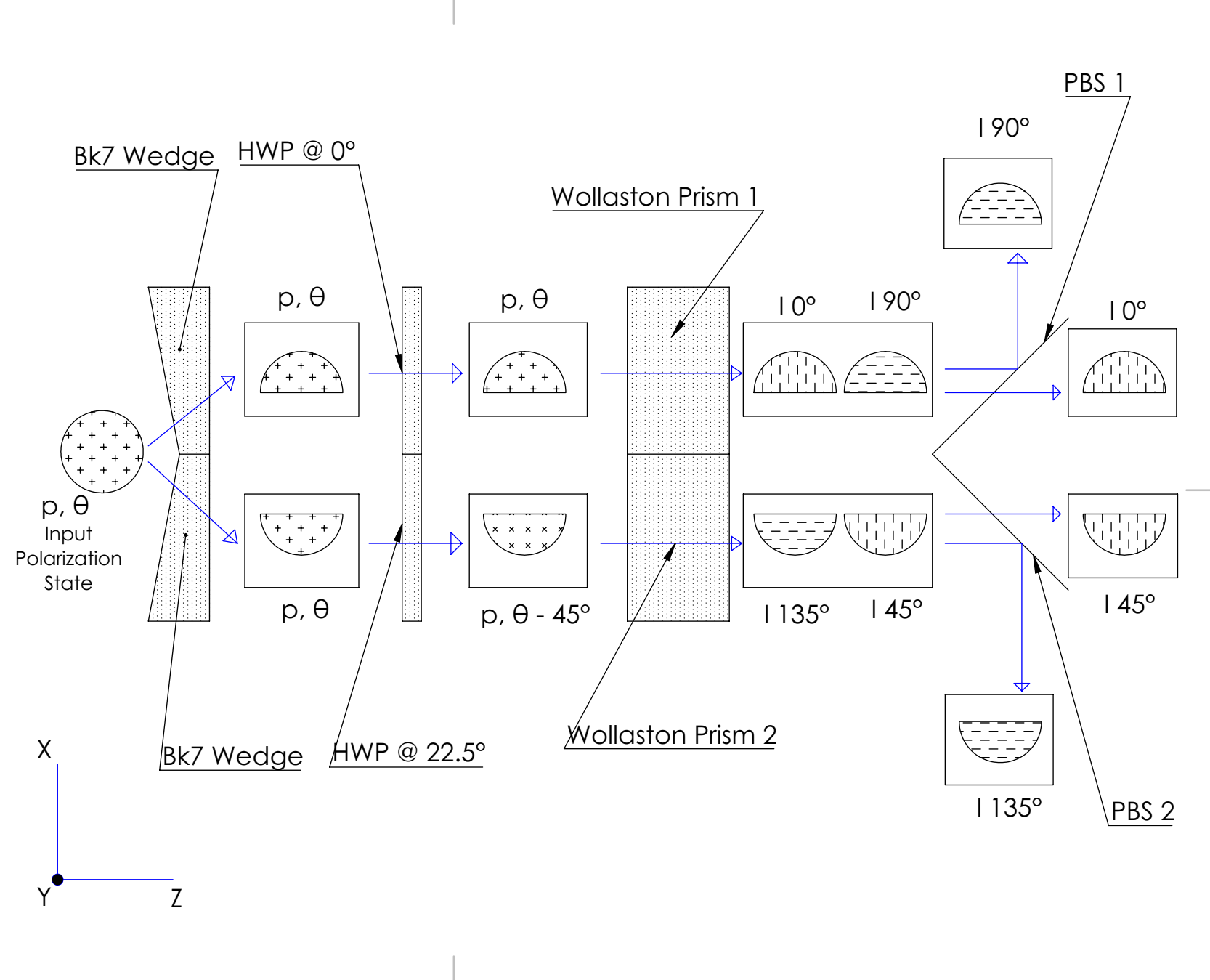}}
    \caption{Schematic of the working of the polarizer assembly of the WALOP-South instrument. In combination, the Wollaston Prism Assembly consisting of the two BK7 glass wedges, Wollaston Prisms (WP) and Half-Wave Plates (HWP) and the two PBSs act as the polarization beamsplitter unit of the instrument. The collimated beam at the pupil is split equally between two BK7 wedges which is then fed to the twin HWP + WP system to be split into four channels with the polarization states of $0^{\circ}$, $45^{\circ}$, $90^{\circ}$ and $135^{\circ}$, and two PBSs steer these four beams in four directions. The change in the polarization state of the beams while passing through this system is annotated.}
    \label{pol_ass_cartoon}
\end{figure}

It consists of four sub-assemblies: (a) Wollaston Prism Assembly (WPA), (b) Wire-Grid Polarization Beam-Splitter (PBS), (c) Dispersion Corrector Prisms (DC Prisms) and (d) Fold Mirrors. The WPA, using the splitting action of the WPs, separates the beam at the pupil into O1, O2, E1 and E2 beams- corresponding to the  polarization angles of $0^{\circ}$, $45^{\circ}$, $90^{\circ}$ and $135^{\circ}$ respectively. The PBS' act as beam selectors, allowing both the O beams to pass through while folding the E1 and E2 beams along -x and +x directions. Fig~\ref{pol_ass_cartoon} shows the overall working idea of the WPA and PBS components of the polarizer assembly, which in combination act as the polarization beamsplitter unit of the instrument. The collimated pupil is split equally between two BK7 wedges which is then fed to the HWP + WP system to split into four channels. Then the two PBS' steer the four beams in four directions. The need for the PBS' is described in Sec~\ref{sec-pbs}. The DC Prisms are a pair of glass prisms present in the path of each of the four beams after the PBS to correct for the spectral dispersion introduced by the WPA. Additionally, before the O-beams enter their respective camera assemblies, mirrors placed at $\pm~45^{\circ}$ to the y-axis in the y-z plane fold the beams into +y and -y directions. This folding was done to limit the length of the instrument to 1.1~m from the telescope focal plane. 

\subsubsection{Wollaston Prism Assembly (WPA) Architecture}\label{WPA Architecture}
\par
\begin{figure}
    \centering
    \includegraphics[scale = 0.5]{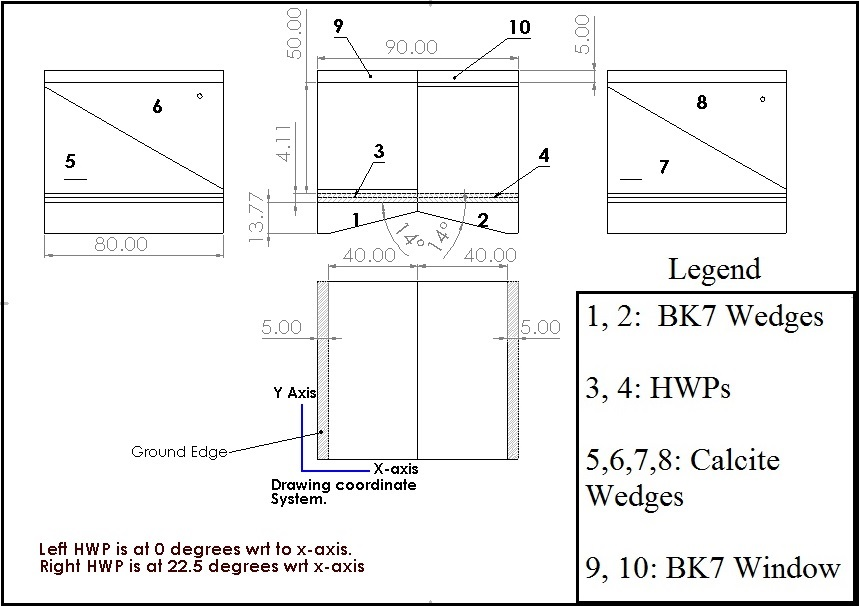}
    \caption{Drawing of the Wollaston Prism Assembly (WPA). The optic axis of the calcite wedges forming the Wollaston Prisms are marked. All length dimensions are in mm.}
    \label{WPA_drawing}
\end{figure}

 Fig~\ref{WPA_drawing} shows the drawing of the WPA.  The WPA is made of two parts, named as the Left Half and the Right Half. Each half has the following optical components: a glass (BK7) wedge, a half-wave retarder plate (HWP), a calcite WP and a flat BK7 window. The aperture of all the elements is $45~mm\times80~mm$, making the overall WPA aperture $90~mm\times80~mm$. The Left Half has been designed to separate the $0^{\circ}$ and $90^{\circ}$ polarizations, from which the Stokes parameter $q$ will be obtained while the Right Half separates the $45^{\circ}$ and $135^{\circ}$ polarizations from which the $u$ parameter will be obtained. For this purpose, the Left HWP's fast-axis is along $0^{\circ}$ and that of the Right one is at $22.5^{\circ}$ with respect to the x-axis in the WPA coordinate system. While the right HWP effectively rotates the EVPA (electric vector polarization angle) of the beam by $-45^{\circ}$, the left HWP has no effect on the beam, and has been used to maintain similar optical path as the right HWP. Both WPs separate $0^{\circ}$ and $90^{\circ}$ polarizations. So while the right WP separates $0^{\circ}$ and $90^{\circ}$ polarizations of the incoming beam, because the HWP has rotated the beam by $-45^{\circ}$, it effectively separates $45^{\circ}$ and $135^{\circ}$ polarizations. At the exit face, a BK7 window is placed to provide a surface which can be anti-reflection (AR) coated. Calcite is a softer material than conventional glasses. During application of AR coatings, large mechanical and thermal stresses that develop on substrates could break the large calcite wedges. The complete WPA is cemented together as one unit using the Norland 65 cement. This is a flexible cement which has been carefully chosen so as to withstand large stresses in WPs that will arise from anisotropic thermal expansion of calcite WPs\cite{Pellicori:70} when subjected to large temperature variations such as that expected at SAAO's Sutherland Observatory (Table~\ref{telescope_details}). To ascertain the flexibility of the cement, sacrificial calcite WPs cemented with Norland 65 were subjected to temperatures in range of $-10^{\circ}~C$ to $40^{\circ}~C$ in an environmental chamber.

\begin{figure}
    \centering
    \frame{\includegraphics[scale = 0.75]{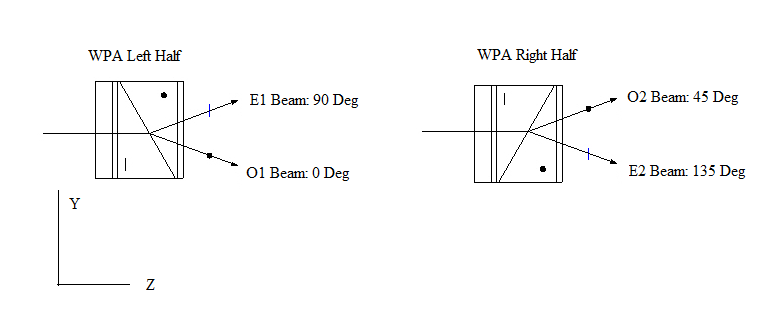}}
    \caption{Beam splitting action of the Left and Right Half of the Wollaston Prism Assembly (WPA) as seen from Y-Z plane. The right WP is rotated by $180^{\circ}$ with respect to the left Wollaston Prism. }
    \label{WP_working}
\end{figure}

\par Due to extended FOV of WALOP-South, rays from non-axial field points arrive at large oblique angles at the pupil. This angle $\alpha$'s approximate value can be found using Eqn~\ref{alpha1} (also called Optical Invariant equation), where $D$, $p$ and $\theta$ are the telescope primary mirror diameter, pupil image diameter and the on-sky angle of the field point from center. Without the BK7 wedge, a fraction of rays from such sources will go on to hit the interface between the Left and Right Half, leading to throughput loss and possible stray light from scattering at the surface. To avoid such a scenario, BK7 wedges with tilt angle of $14^{\circ}$ are used. Fig~\ref{WALOP-S-Ebeam} shows the BK7 wedges bending the rays from the whole field such that none hit the interface. 

\begin{equation}\label{alpha1}
    \alpha = \frac{ D \times\theta}{p}
\end{equation}

\par The HWPs are made of Quartz-MgF$_{2}$ plates designed to provide achromatic half-wave retardation (within 0.515 to 0.480 $\lambda$) over the wavelength range of $0.5~{\mu}m$ to $0.7~{\mu}m$ for normal incidence. Both the Left and Right HWPs are made by a $2\times1$ mosaic of $45~mm\times40~mm$ size HWPs. This was necessitated due to unavailability of larger blanks of Quartz and MgF$_{2}$ to create a single HWP of aperture $45~mm\times80~mm$. In general, the retardance  of a HWP is dependant on the absolute and azimuth angles of incidence of the incoming beam with respect to the fast axis of the HWP\cite{HWP_retardance}. Due to the bending by the BK7 wedges, rays from the entire FOV have oblique angles of incidence on both the HWPs, and hence undergo 'non-half wave' retardation. This leads to cross-talk between all the Stokes parameters, which has been considered and modelled as part of the calibration method we have developed for the instrument (Section~\ref{calibration}).

\par Two identical calcite WPs are used as polarization analyzers. The polarization separation behaviour of the WPs is shown in Figure~\ref{WP_working}. While both the WPs separate $0^{\circ}$ and $90^{\circ}$ (called O and E beams) polarizations, due to the HWPs in front of each WP, the O1 beam and O2 beams correspond to  $0^{\circ}$ and $45^{\circ}$ while the E1 and E2 to $90^{\circ}$ and $135^{\circ}$. The right WP is rotated by $180^{\circ}$ with respect to the left WP. Hence, the O1 beam goes downwards and O2 beam goes upwards, and vice versa for E1 and E2 beams. This inversion of WPs was done so that the E1 and E2 and O1 and O2 beams can be folded in opposite directions to each other to physically separate the four beams to correct for their dispersion, the need for which is explained in Section~\ref{dispersion}.

\par The complete polarizer assembly is fabricated by \href{http://www.klccgo.com/}{Karl Lambrecht Corporation}, Chicago, USA.

\subsubsection{Choice of Wollaston Prism (WP)}\label{WP_design}

A WP can be characterized by the following quantities: (a) aperture, (b) material, (c) wedge angle, (d) optic axis directions in the two wedges forming the WP and (e) extinction ratio. Of these, the material and the wedge angle decide the split angle of the WP.

The minimum required split angle for a WP ($\beta_{min}$) to separate the E and O beams such that they are fully separated on the detectors is given by Eqn~\ref{beta_not}. It is twice of the angle formed by the extreme field point in the direction of splitting at the pupil (which is y-axis in our case), given by Eqn~\ref{alpha1}. For a FOV of $34.8\times34.8$~arcminute ($0.58^{\circ}\times0.58^{\circ}$), the extreme vertical field point is $0.29^{\circ}$ for which $\beta_{min}$ = $8.92^{\circ}$.

\begin{equation}\label{beta_not}
    \beta_{min} = \frac{2\times D \times\theta}{p}
\end{equation}

Materials from which WPs of such large split angle can be created in the optical wavelengths are $\alpha$-BBO (Barium Borate), LiNbO$_{3}$ (Lithium Niobate) and calcite. $\alpha$-BBO and LiNbO$_{3}$ have lower birefringence than calcite and are not available in large crystal sizes. Employing WPs made from either of these would lead to reduction in the pupil image size and consequently higher split angle requirement (Eqn~\ref{alpha1}) and hence larger dispersion and aberrations. Also, calcite has been used in previous wide-field astronomical polarimeters such as SALT-RSS polarimeter\cite{salt_commisioning}. Calcite WPs have very high extinction ratios $>10^{5}$.

The approximate split angle ($\beta$) for a WP can be calculated using Eqn~\ref{split_angle}  \cite{Simon:86}, where $n_{e}$ and $n_{o}$ are extraordinary and ordinary index of refraction and $\gamma$ is the wedge angle of the WP.
\begin{equation}\label{split_angle}
    sin({\frac{\beta}{2}}) = (n_{e} - n_{o})tan{\gamma}
\end{equation}

The WP used in the WPA has a wedge angle of $\gamma = 30^{\circ}$, for which the split angle comes out to be $11.4^{\circ}$ at $0.6~{\mu}m$ (Figure~\ref{WP_dispersion}). As can be seen, the split angle is larger than $\beta_{min}$. We increased the FOV from $30.0~\times~30.0$~arcminutes to  $34.8~\times~34.8$~arcminutes to take advantage of this large split angle without degrading the image quality. The plate scale for this FOV is 0.5"/pixel, so the median seeing FWHM (1.5") is sampled by 3 pixels. Although we could have increased the FOV further, it would lead to sparser PSF sampling at the detector and higher instrumental noise due to pixel to pixel sensitivity variation of the CCD.

As can be seen from Eqn~\ref{beta_not}, for the same telescope, if the pupil diameter is larger, WP with smaller split angle is required which will introduce smaller dispersion. The issue of dispersion from calcite WPs is described in Section~\ref{dispersion}. So, while it is advantageous to make the pupil larger to reduce dispersion by using large aperture WPs, the upper limit is set by the availability of largest calcite crystals. Additionally, making a large pupil makes the collimator design challenging. Hence we decided to use WPs with the largest available apertures- $45~mm\times80~mm$.

\subsubsection{Dispersion by WP and the WPA} \label{dispersion}
\begin{figure}
    \frame{\includegraphics[scale = 0.4]{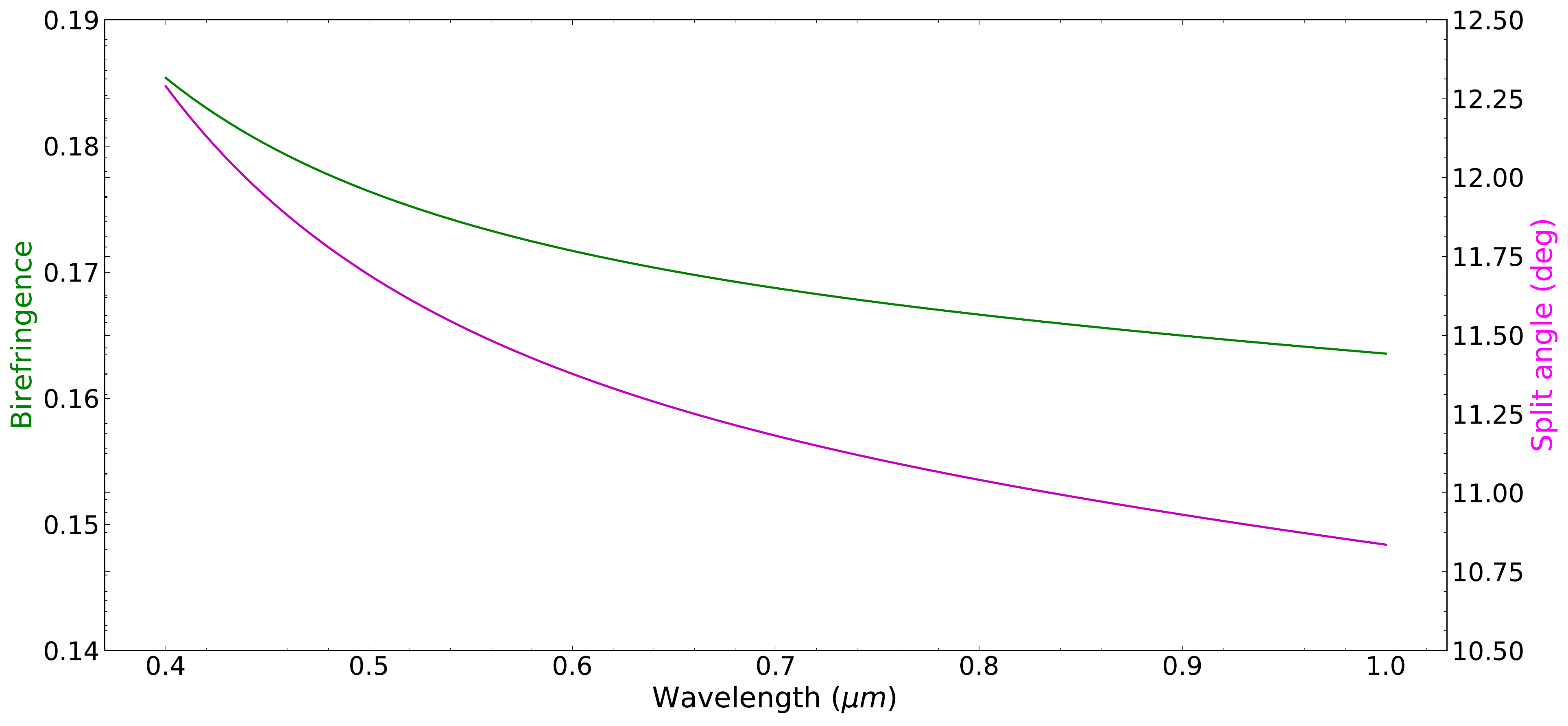}}
    \caption{Dependence of birefringence and split angle of the WALOP Wollaston Prisms on wavelength.}
    \label{WP_dispersion}
\end{figure}

\begin{figure}
    \centering
    \includegraphics[scale=0.35]{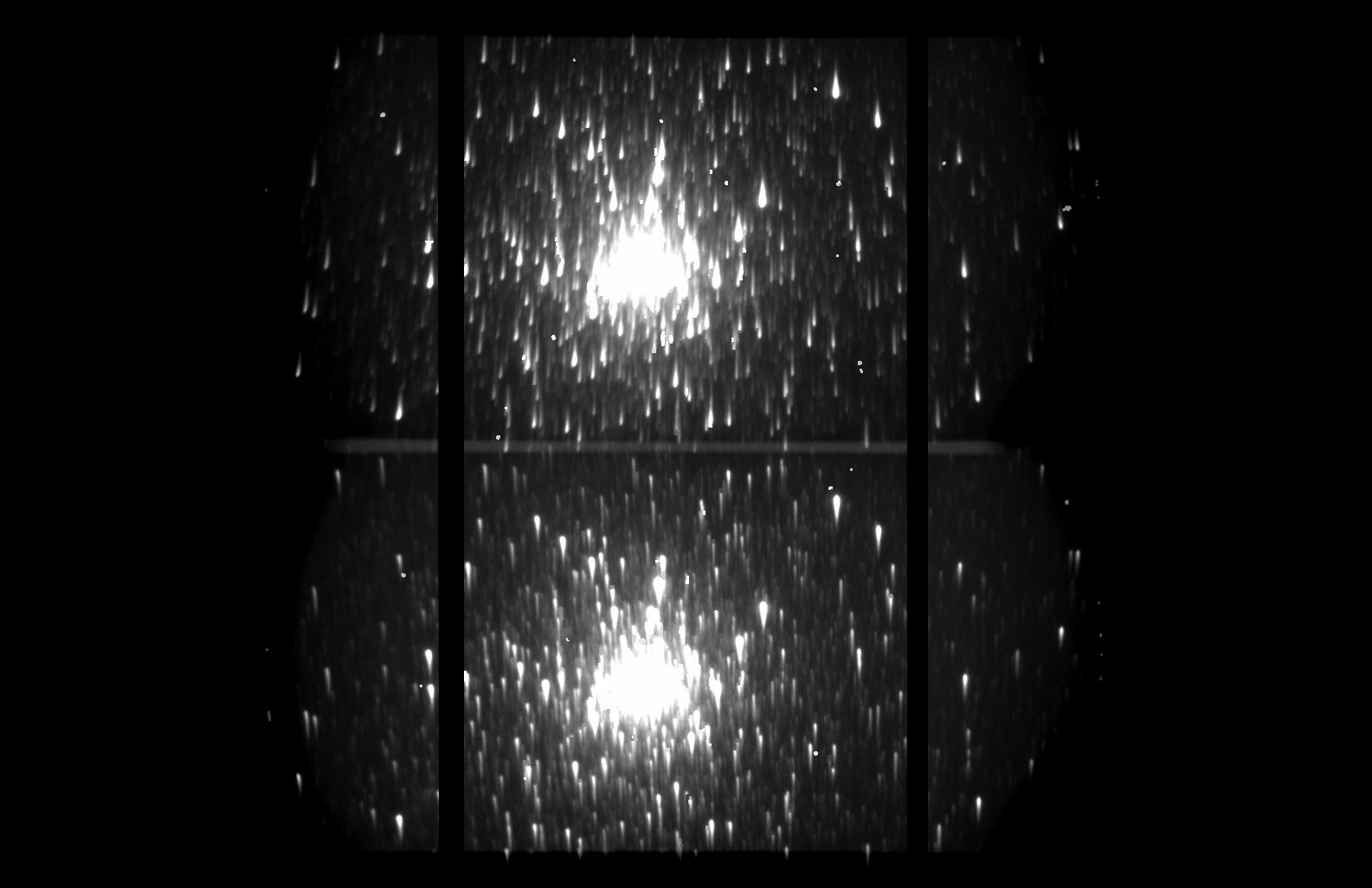}
    \caption{ SALT-RSS two channel polarimeter image of the M30 globular cluster, which employs calcite Wollaston Prisms as the polarization analyzer. The E and O beams are imaged on the top and bottom half of the detector. The dispersion in the E and O images for a $4\times8$~arcminute field is clearly visible. Also, the dispersion direction is opposite in the two images. This image is reproduced with permission, courtesy of Prof. Kenneth Nordsieck, the PI of the RSS instrument.}
    \label{salt_dispersion}
\end{figure}

Birefringence ($\delta{n} = n_{e} - n_{o} $) of crystals and consequently the split angle of WPs depends on the wavelength (Eqn~\ref{split_angle}). Figure~\ref{WP_dispersion} shows the wavelength dependence of calcite's birefringence and split angle for the WALOP WPs. This split angle variation with wavelength leads to the outgoing E and O beams to have dispersion in the splitting direction. WPs with larger split angle introduce larger dispersion. There is significant change in the split angle within the SDSS-r band wavelengths of $0.55{\mu}m$ to $0.7{\mu}m$. In case of WALOP WPs, this leads to dispersion in the y-axis direction in the outgoing E and O beams corresponding to a spectral resolution $R \sim 80$. The dispersion in the E and O beams are similar (but not equal) and in opposite directions. Hence the dispersion of both the E and O beams cannot be corrected by placing a suitable glass wedge after the WP exit surface. Fig~\ref{salt_dispersion} shows the dispersion in the E and O images for a $4\times8~arcminute$ FOV of the SALT-RSS polarimeter which employs calcite WPs as the polarization analyzer. As can be seen, the dispersion of the same stars is in the opposite directions between the top and bottom images. In addition to the WPs, the BK7 wedges at WPA entrance also introduce dispersion, but in the x-axis (horizontal). So, the outgoing beams from the WPA suffer from dispersion along both the x and y directions.

\par Most imaging polarimeters like RoboPol which do not image the entire FOV on separate detectors use WPs with small split angles (around ~$1.0^{\circ}$), for which the dispersion is minimal and lower in value than other optical aberrations of the system. On the other hand, SALT-RSS (Fig~\ref{salt_dispersion}) and HOWPol polarimeters image a large FOV on separate detectors, but do not correct for the dispersion by the WPs and are limited to carry out only narrow-band imaging polarimetry.

\subsubsection{Wire-Grid Polarization Beam-Splitter (PBS)}\label{sec-pbs}
As the dispersion due to the WPs in E and O beams is in opposite directions, the four beams need to be separated physically so that the required correction can be applied on each beam. The wire-grid PBS is designed such that it allows the O-Beam to pass through but folds the E-Beam by acting as a mirror. Since the two beams have already been completely separated in the angle-space by the WPA, the lower extinction ratio of the PBS for the E-Beam (which causes some fraction of O-beams to be folded as well, but at different angles than E-beam) will not lead to any O-Beam rays reaching the detector of E-Beam cameras. 
Unlike other types of polarization beam splitter systems, wire-grid type have uniform performance for large angles of incidence of $\pm~10^{\circ}$ and are available in  sizes required for WALOP-South instrument. The PBSs used in the WALOP-South design are of aperture $78\times90~mm$, and have been fabricated by \href{https://moxtek.com/}{Moxtek, Inc.}, Utah, USA. It has nearly 90~\% throughput for p (O Beam) and s (E Beam) polarizations for the transmitted and reflected beams respectively over the SDSS-r filter wavelengths.

\subsubsection{Dispersion Corrector Prisms (DC Prisms)}\label{WP_disp}

Each of the four beams have two DC Prisms- the first prism made of S-TIH6 glass corrects the y direction dispersion and the second prism made of S-BSL7 (glass from Ohara catalogue with properties identical to BK7) corrects the x direction dispersion. Each prism has one tilted and one flat surface, as shown in Fig~\ref{WALOP-S-Obeam1}, \ref{WALOP-S-Obeam2} and \ref{WALOP-S-Ebeam}. Details of the DC Prisms are captured in Table~\ref{dc_prisms}. 

\begin{table}
    \centering
    \begin{tabular}{|c|c|c|c|c|}
        \hline
        \textbf{DC Prism No}. & \textbf{Beam} & \textbf{Glass Name} & \textbf{Tilt Angle-X} & \textbf{Tilt Angle-Y} \\
         \hline
        Prism 1 & O Beam &  S-TIH6 & - & $8.3^{\circ}$ \\
         \hline
        Prism 1 & E Beam &  S-TIH6 & - & $7.76^{\circ}$ \\
         \hline
        Prism 2 & O Beam &  S-BSL7 & $14.0^{\circ}$ & - \\
         \hline
        Prism 2 & E Beam &  S-BSL7 & $14.0^{\circ}$ & - \\
         \hline
    \end{tabular}
    \caption{Details of the Dispersion Corrector Prisms used in WALOP-South optical model.}
    \label{dc_prisms}
\end{table}

\subsection{Collimator Assembly}\label{collimator}
 The collimator assembly begins from the telescope focal plane as shown in Fig \ref{WALOP-S}, \ref{WALOP-S-Obeam1}, \ref{WALOP-S-Obeam2} and \ref{WALOP-S-Ebeam}. It consists of 6 spherical lenses placed axially along the z-axis. At the telescope focus, the field size is $162~mm\times162~mm$, from which a pupil of of $65~ mm$ in diameter is formed. The choice of the pupil size was driven by the architecture of the polarizer assembly described in Section~\ref{WP_design}. The effective focal length of the collimator is $1.6\times10^{8}$~mm and its total length is 700~mm.

\subsection{Camera Assemblies}\label{camera}
\begin{figure}
    \centering
    \fbox{\includegraphics[scale=0.15]{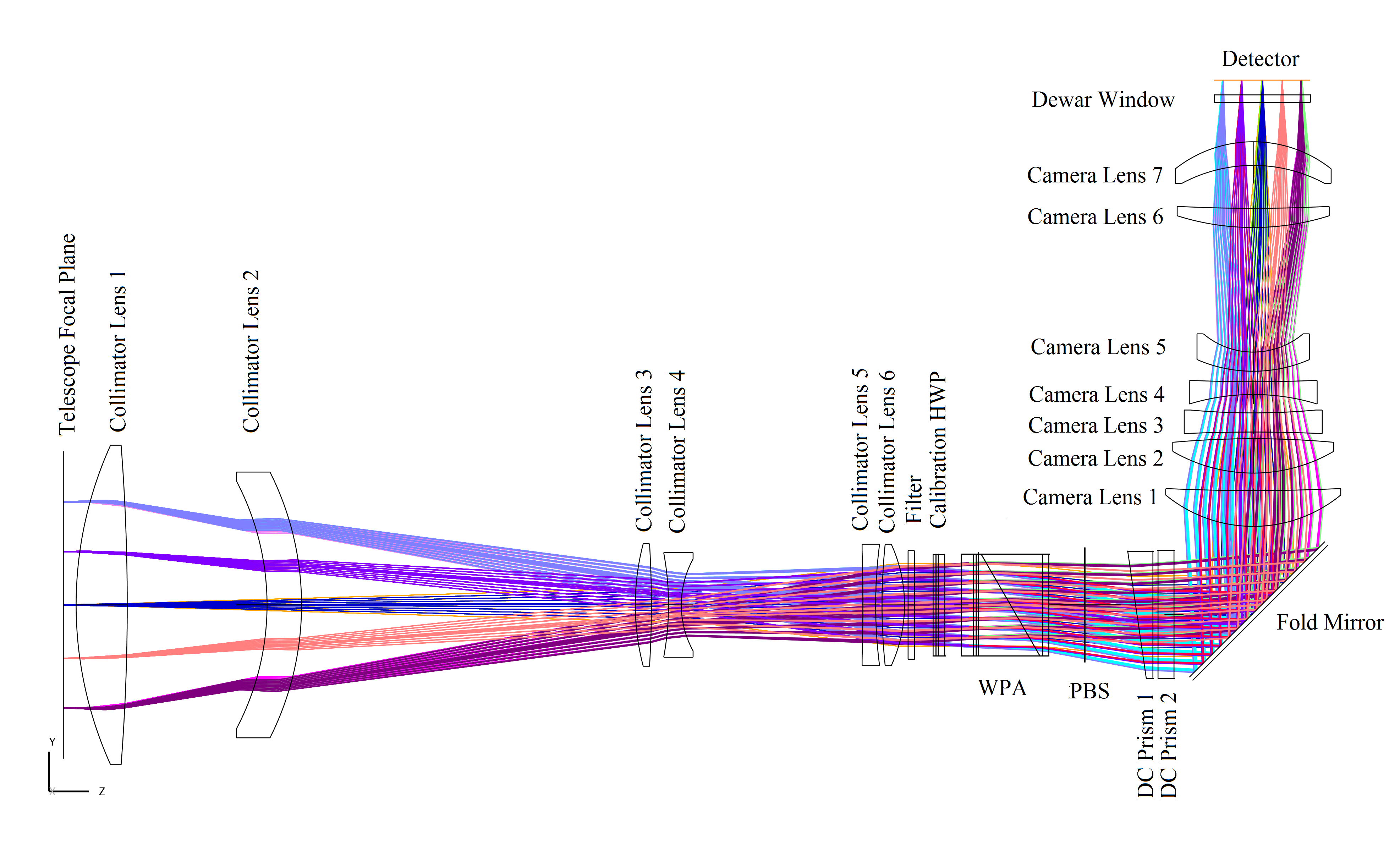}}
    \caption{O1 Beam Optical Path. After getting split by the Wollaston Prism Assembly (WPA), the Polarization Beam-Splitter (PBS) allows this beam to pass through it to the Dispersion Corrector (DC) Prisms. The fold mirror downstream then directs this beam along +y direction to its camera.}
    \label{WALOP-S-Obeam1}
\end{figure}

\begin{figure}
    \centering
    \fbox{\includegraphics[scale=0.15]{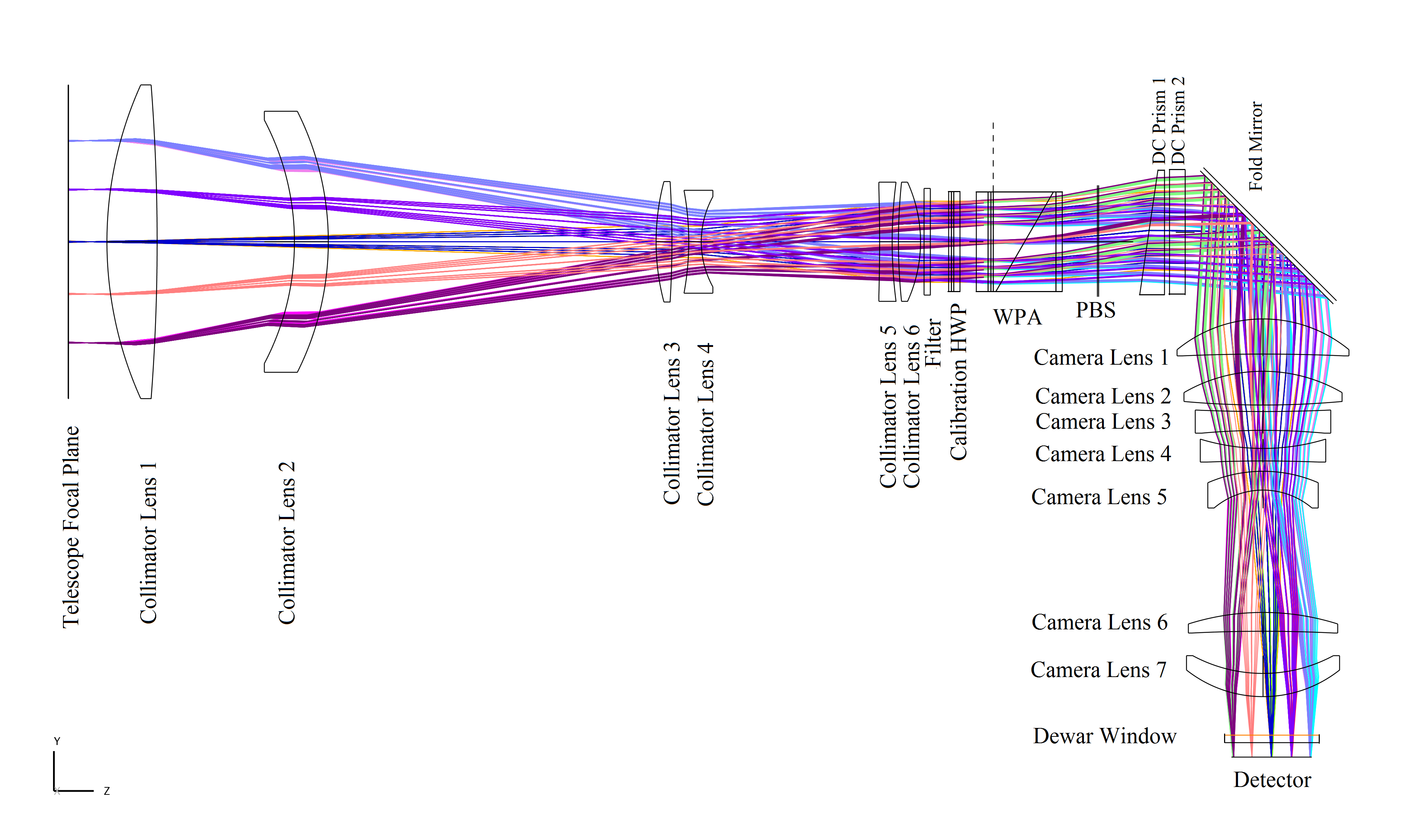}}
    \caption{O2 Beam Optical Path. After getting split by the Wollaston Prism Assembly (WPA), the Polarization Beam-Splitter (PBS) allows this beam to pass through it to the Dispersion Corrector (DC) Prisms. The fold mirror downstream then directs this beam along -y direction to its camera.}
    \label{WALOP-S-Obeam2}
\end{figure}

\begin{figure}
    \centering
    \fbox{\includegraphics[scale=0.2]{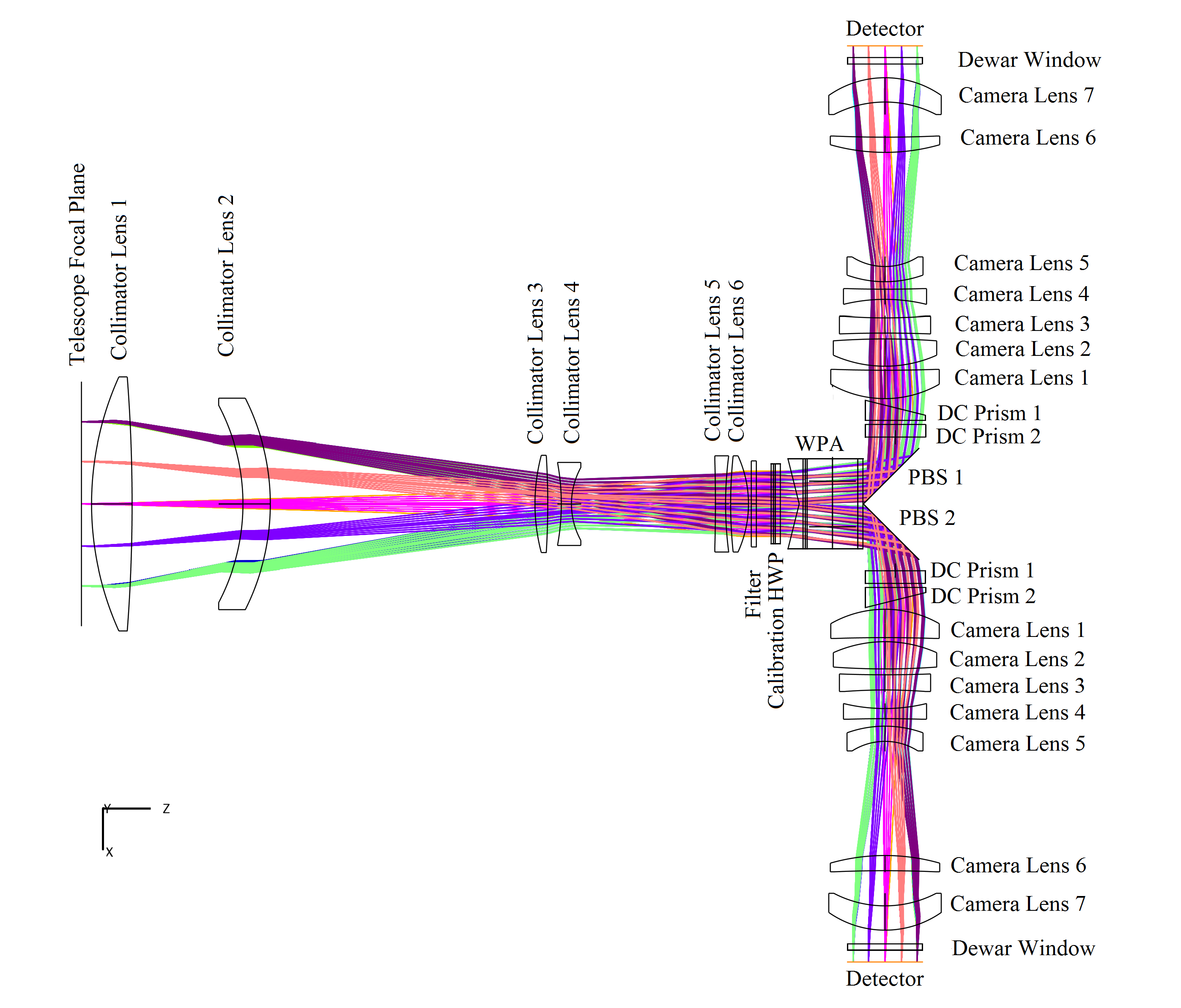}}
    \caption{E1 and E2 Beam Optical Paths. After getting split by the Wollaston Prism Assembly (WPA), the two Polarization Beam-Splitters (PBS) fold these along -x and +x directions. After passing through the Dispersion Corrector (DC) Prisms, each beam goes to its camera.}
    \label{WALOP-S-Ebeam}
\end{figure}

Each camera assembly consists of 7 spherical lenses. Fig~\ref{WALOP-S-Obeam1} and \ref{WALOP-S-Obeam2} shows the optical path for O1 and O2 beams. The Polarization Beam-Splitter (PBS), described in Section~\ref{pol_design}, allows these beams to pass straight through to be then folded towards the +y and -y directions by the Fold Mirrors.  Fig~\ref{WALOP-S-Ebeam} shows the optical system for the E1 and E2 beams. The PBS folds these beams towards the -x and +x directions. 
\par The O1 and O2 cameras are identical to each other, i.e, they have identical lens prescription and air gaps. Similarly, the E1 and E2 cameras are identical to each other. Between the O and E beam cameras, while the glass types are same for corresponding lenses, their apertures and air separations are different. Also, between the O-Beam and E-Beam cameras, the radii of curvatures and central thicknesses are same for the first five lenses but differ for the last two lenses. As a design choice, we made the E and O beam cameras as similar as possible to reduce fabrication effort and cost of the lenses. The f-number at the detectors is 6.12 and 6.06 for the E and O beam cameras, respectively, which corresponds to effective focal lengths of 6220~mm and 5993~mm. With $4k\times4k$ detectors of pixel size $15~{\mu}m$, the plate scale is 0.5 arcsecond per pixel. This allows for better than Nyquist sampling for the median seeing FWHM of 1.5 arcsecond at the Sutherland Observatory.

\subsection{Filter, Calibration HWP}
\begin{figure}
    \centering
    \includegraphics[scale = 0.4]{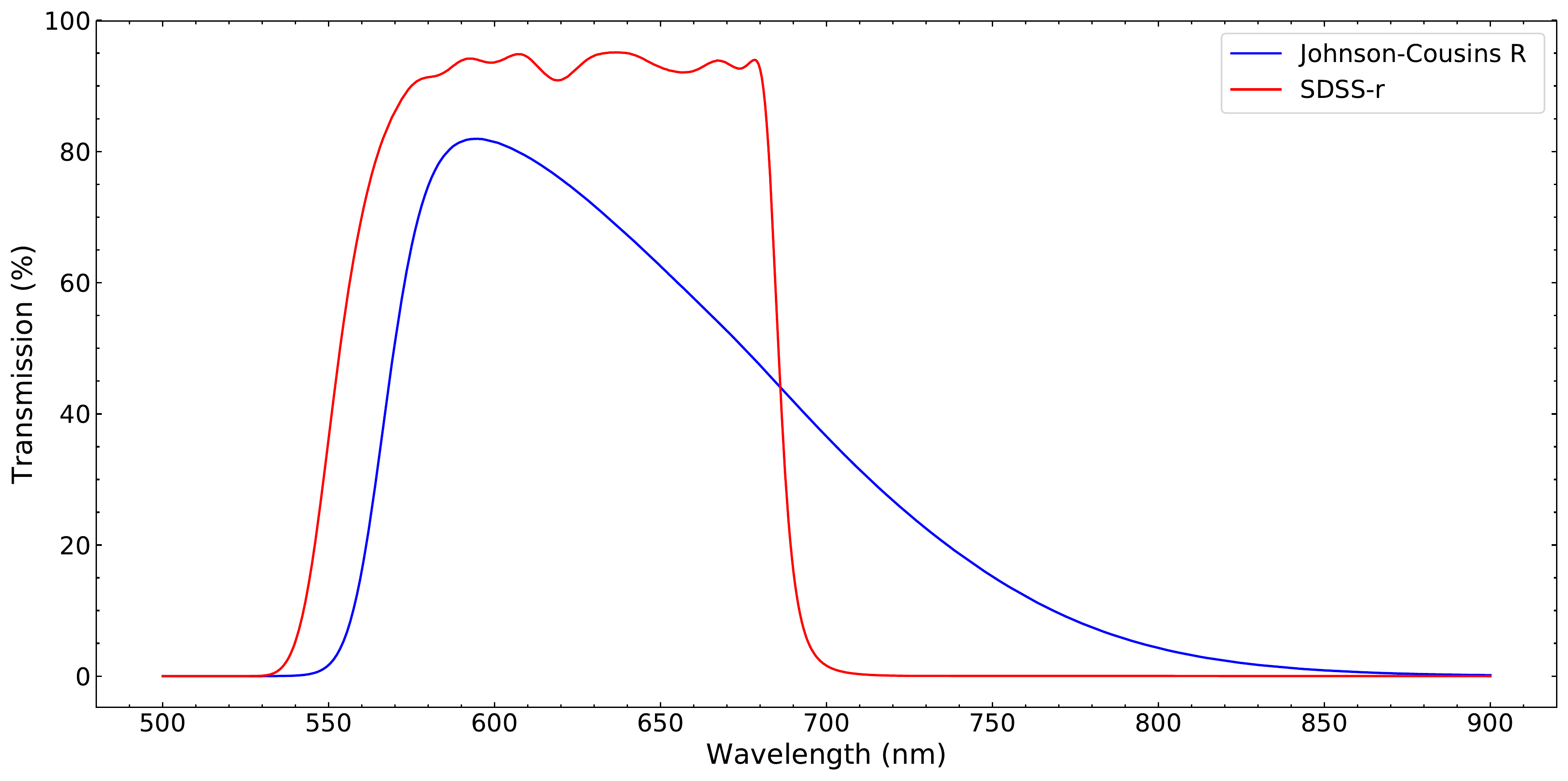}
    \caption{Comparison of the transmission profiles of the SDSS and Johnson-Cousins R filters. While both filters have similar throughput, the SDSS-r has a narrower wavelength coverage which reduces the dispersion by the Wollaston Prisms.  }
    \label{filter_comparison}
\end{figure}

The filter and the calibration HWP are located near the pupil image before the polarizer assembly, as shown in Fig~\ref{WALOP-S-Obeam1}, \ref{WALOP-S-Obeam2} and \ref{WALOP-S-Ebeam}.
The initial design idea was to use Johnson-Cousins R as the main broadband filter. In comparison to this filter, SDSS-r has a similar throughput but a smaller wavelength spread which helps in reducing the spectral dispersion introduced by the polarizer assembly and provides better PSFs (refer to Section~\ref{performance}, Fig~\ref{PSF_Johnsons_SDSS}). Fig~\ref{filter_comparison} compares the transmission profiles between the two filters. The SDSS-r filter used in WALOP-South instrument is of $85~mm$ aperture fabricated by \href{https://www.asahi-spectra.com/}{Asahi Spectra Co., Ltd. (ASC), Japan}.
\par Rotation of the calibration HWP to orientations that are multiples of $45^{\circ}$ deg, would in an ideal instrument interchange the intensities in the O and E beams. Any departure from this behaviour of the beams passing through the four cameras can be used to correct for any channel dependant transmission variation. This offers an additional method to calibrate the instrument (Section~\ref{calibration}) during main science observations. The aperture of the calibration HWP is $80~mm\times80~mm$, and is made by a $2\times2$ mosaic of $40~mm\times40~mm$ size HWPs whose fast-axes are accurately aligned with each other.

\section{Optical Performance}\label{performance}
\par
\begin{figure}[hbt!]
\centering
\includegraphics[scale = 0.5]{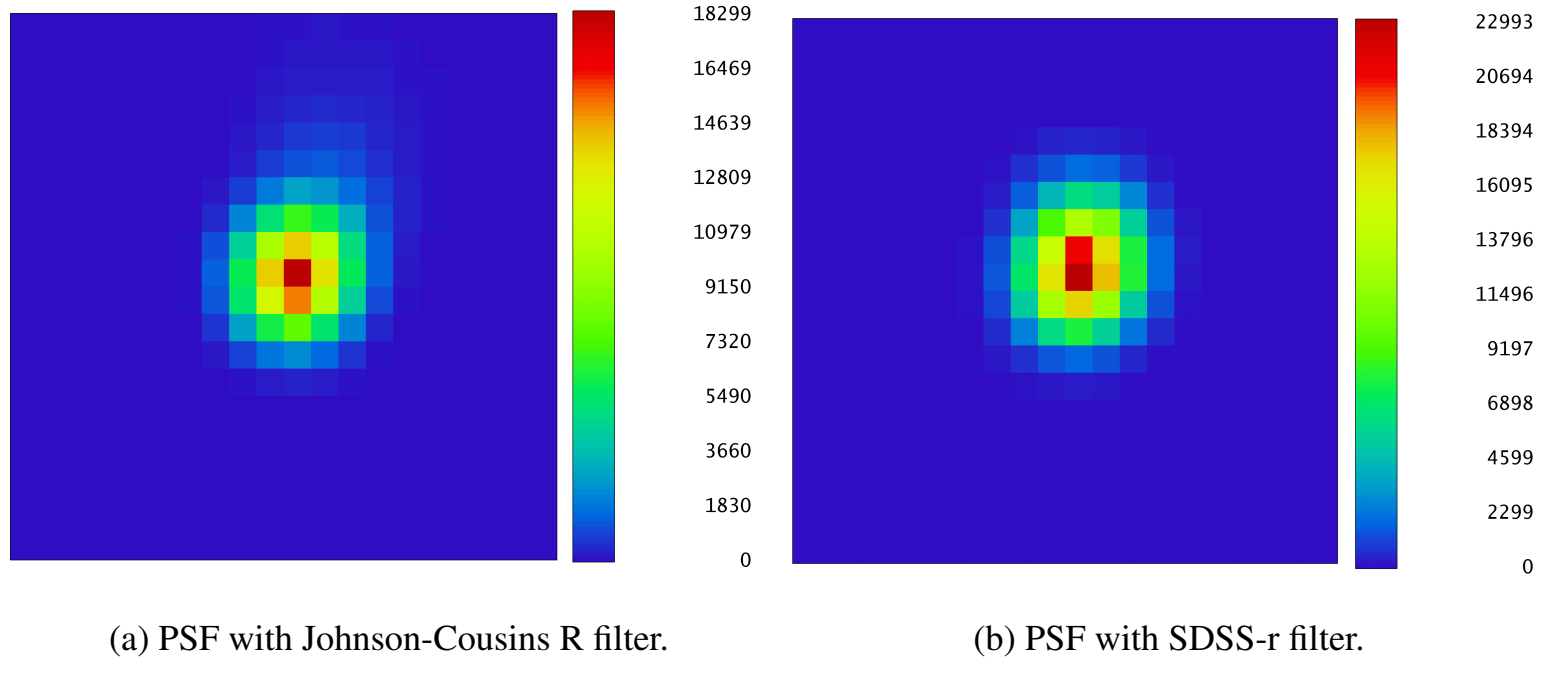}
\caption{Comparison of the PSF at the center of FOV for a 1.5" (median seeing) input beam at the detector for Johnson-Cousins and SDSS R band filters. As can be seen, with Johnson-Cousins filter, we get an elongated PSF due to the residual dispersion from the Wollaston Prisms owing to the filter's broad spectral range while we get a symmetric PSF with the SDSS filter.}
\label{PSF_Johnsons_SDSS}
\end{figure}

 As described in Section~\ref{techgoals}, the essential performance criterion for the instrument optical design is to obtain near seeing limited PSF for the entire FOV (within 1.5-2 times the median seeing Gaussian profile). To carry out polarimetry with sensitivity of 0.05~\%, all the photons from a star reaching the detector need to be counted using aperture photometry. Larger PSFs sample larger sky areas and lead to higher measurement uncertainty. So, in judging the quality of PSFs obtained, while FWHMs are a good measure of spatial resolution, for high sensitivity polarimetry, the ensquared energy radius containing all the starlight ($>99.95~\%$) is a better measure. For example, Fig~\ref{PSF_Johnsons_SDSS} shows the PSF at the center of WALOP-South's FOV for the Johnson-Cousins R and SDSS-r filters. Although both have similar FWHMs, the radius corresponding to the $>99.95~\%$ ensquared energy is much larger for the Johnson-Cousins filter due to its extended spectral transmission tail, as shown in Fig~\ref{filter_comparison}. 

\begin{figure}
    \centering
    \frame{\includegraphics[scale = 0.15]{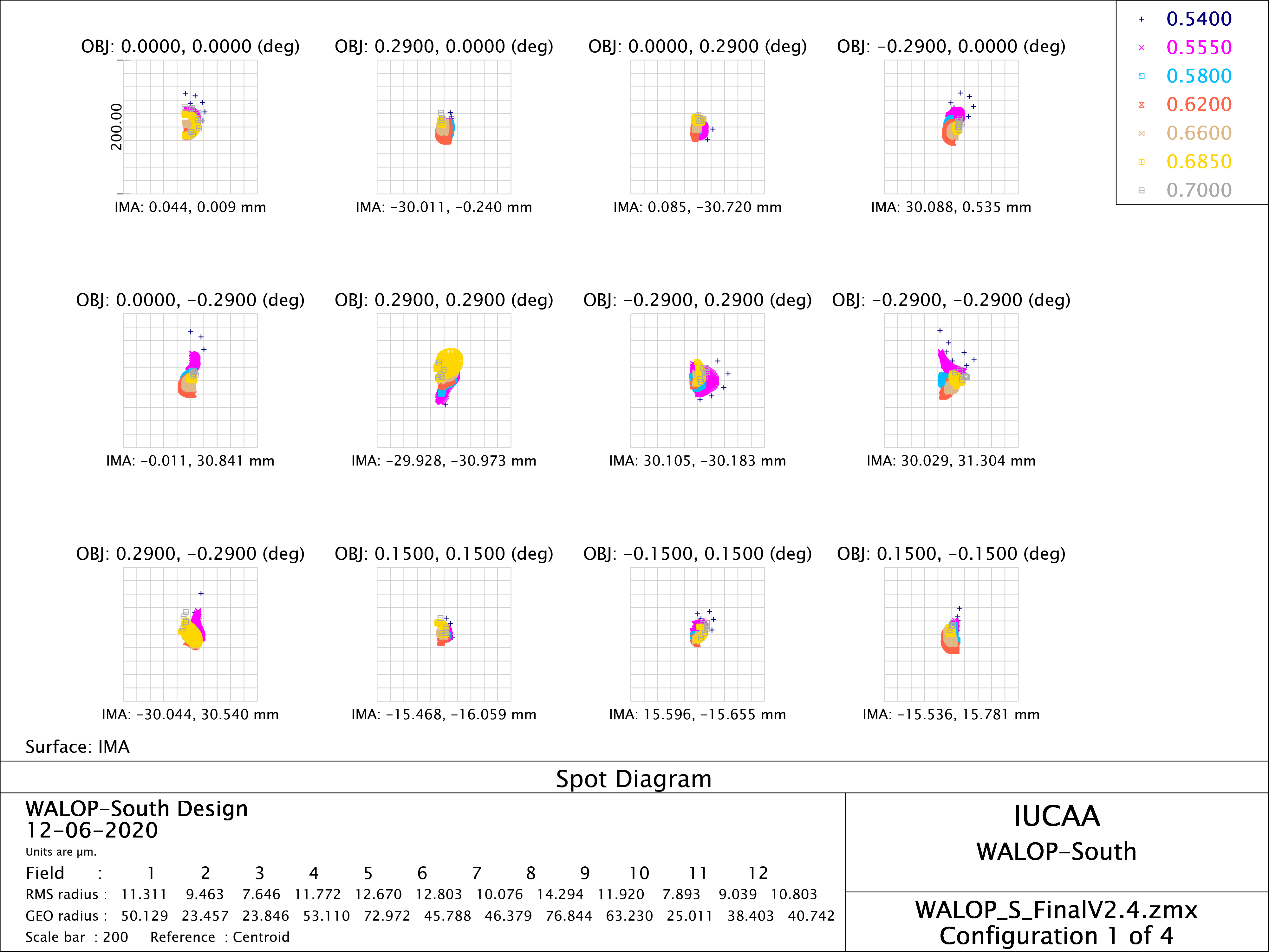}}
    \caption{Spot diagram for the O Beam cameras at the detector for different field points. Different colors represent rays of different wavelengths, as labeled in the image legend. RMS and GEO radius stand for the root-mean square and geometric radius of the spot diagrams, respectively. The optical performance of O1 and O2 beams are identical as they follow identical optical paths.}
    \label{spot_beam1}
\end{figure}

\begin{figure}
    \centering
    \frame{\includegraphics[scale = 0.15]{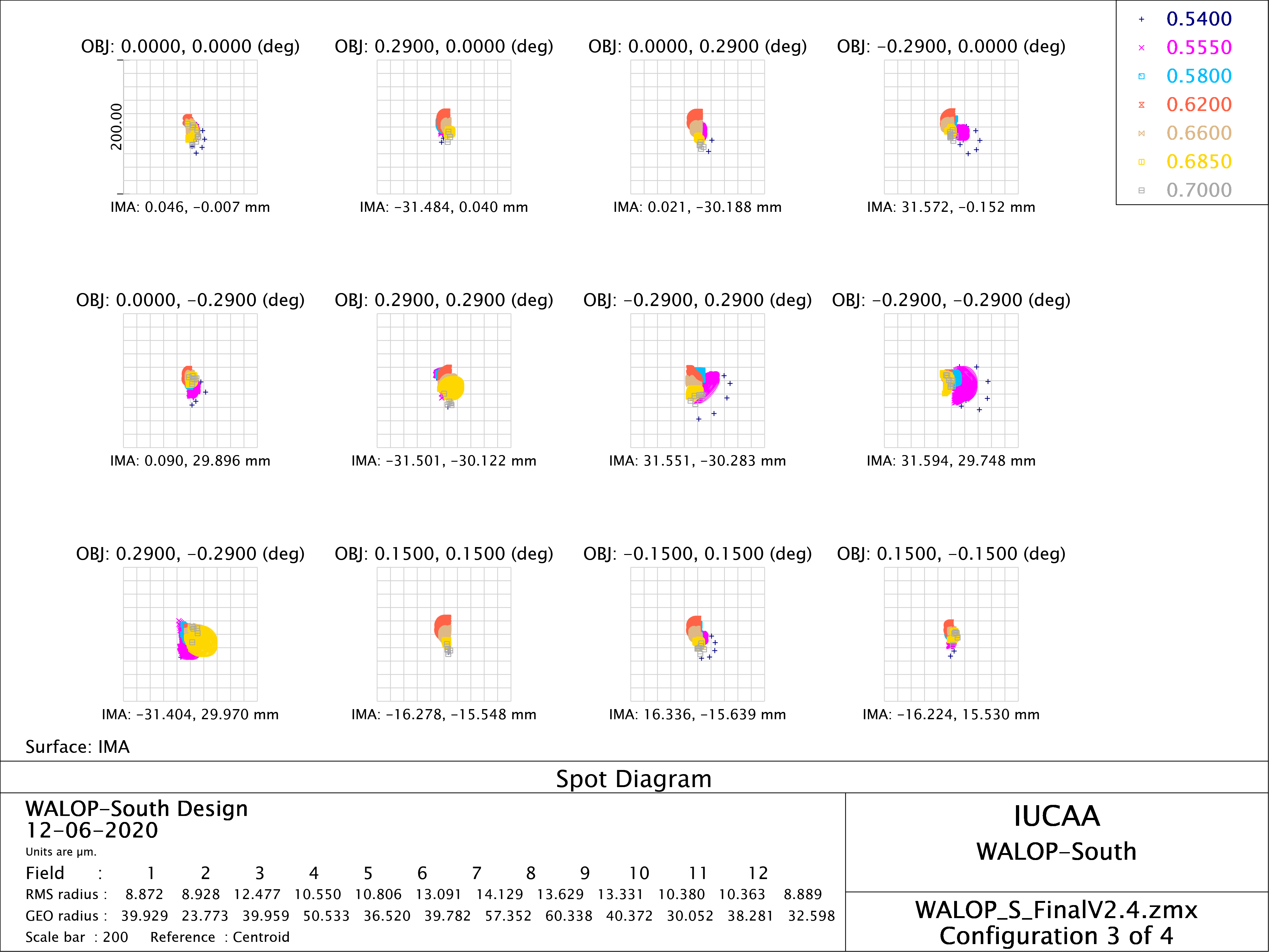}}
    \caption{Spot diagram for the E Beam cameras at the detector for different field points. Different colors represent rays of different wavelengths, as labeled in the image legend. RMS and GEO radius stand for the root-mean square and geometric radius of the spot diagrams, respectively. The optical performance of E1 and E2 beams are identical as they follow identical optical paths.}
    \label{spot_beam3}
\end{figure}

Figs~\ref{spot_beam1} and \ref{spot_beam3} show the spot diagrams for the O and E beam cameras at different field points. As the optical path of the O1 and O2 beams is same, they have identical optical performance. Same is true of E1 and E2 beams. As can be seen, the dispersion introduced by the WPA has been corrected by post-WPA optics. On an average, the centroid of the spot diagrams for all the individual wavelengths within the SDSS-r filter fall within $10~{\mu}m$ (less than a pixel) of each other for the entire FOV for both E and O beams. The average RMS radius of both the beams is around 11.5~${\mu}m$ for the whole field. For comparison, the RMS radius for a 1.5 arcsecond FWHM Gaussian beam (median seeing at Sutherland) with the same plate scale as at the camera detectors is 19.1~${\mu}m$. The PSF obtained will be a convolution of the input Gaussian seeing beam with the spot diagram profile. The radius corresponding to 99.95\% of the ensquared energy at different points in the FOV for a 1.5" FWHM source for the E and O beams is captured in Table~\ref{radii_optical_performance} under column name \textit{Nominal}. The averaged radius for both the beams is less than 75~${\mu}m$ or 5 pixels at the detector whereas for an ideal Gaussian this radius is 67.1~${\mu}m$ or 4.5 pixels at the detector.

\par Tolerance analysis of the optical system was carried out to get an estimate of expected degradation of spot sizes and ensquared energy radii due to tolerances in fabrication of optical components and mechanical mounts to hold the optics. Two compensators were defined - one is the separation between the primary and secondary mirrors of the telescope. The second compensator is the distance of the detector from the last lens in each camera. We used the Monte Carlo (MC) simulation feature of Zemax for this work. Table~\ref{MC_sim} shows the results for the spot radius based on  20,000 Monte Carlo runs for the system. Table~\ref{radii_optical_performance} lists the radius containing
99.95~\% of the ensquared energy  for median seeing conditions for the E and O beams in nominal as well as best and worst Monte Carlo
simulation results at different field points. Even in the worst case scenarios, we get PSFs
equivalent to Gaussian beams with FWHM of 1.6 and 1.3 times the seeing FWHM for the O and E beams, respectively.

\begin{table}[ht!]
\centering
\begin{tabular}{|c|c|c|}
\hline
Parameter                & O-Beams             & E-Beams             \\ \hline
                         & RMS Spot Radius (${\mu}m$) & RMS Spot Radius (${\mu}m$) \\ \hline
Nominal Spot             & 11.63               & 11.77               \\ \hline
Root-Sum-Square          & 17.1                & 15.37               \\ \hline
MC Simulation Best Case  & 11.72               & 11.7                \\ \hline
MC Simulation Worst Case & 37.4                & 25.5                \\ \hline
MC Simulation Mean       & 17.54               & 15.72               \\ \hline
MC Simulation Std Dev    & 0.003               & 0.0018              \\ \hline
\end{tabular}
\caption{Results of Monte Carlo simulations based tolerance analysis for the O and E beams. Root-Sum-Square radius is the RMS spot radius obtained if the offset in spot radius due to all mechanical and optical tolerances are added in quadrature.}
\label{MC_sim}
\end{table}

\begin{table}[ht!]
\begin{tabular}{|c|c|c|c|c|c|c|c|c|}
\hline
\multirow{3}{*}{Sl.   No} & \multicolumn{2}{c|}{Field   Poisition} & \multicolumn{3}{c|}{O   Beams}                                                                       & \multicolumn{3}{c|}{E   Beams}                                                                         \\ \cline{2-9} 
                          & X                  & Y                 & \multicolumn{3}{c|}{\begin{tabular}[c]{@{}l@{}}99.95 \% Ensqaured\\ Energy Radius (${\mu}m$)\end{tabular}} & \multicolumn{3}{c|}{\begin{tabular}[c]{@{}l@{}}99.95 \% Ensqaured \\ Energy  Radius (${\mu}m$)\end{tabular}} \\ \cline{2-9} 
                          & (in Deg)           & (in Deg)          & Nominal                          & MC Best                         & MC Worst                        & Nominal                          & MC Best                          & MC Worst                         \\ \hline
1                         & 0                  & 0                 & 73                               & 75                              & 62                              & 69                               & 69                               & 75                               \\ \hline
2                         & 0.29               & 0                 & 62                               & 61                              & 68                              & 64                               & 63                               & 68                               \\ \hline
3                         & 0                  & 0.29              & 74                               & 72                              & 96                              & 79                               & 82                               & 62                               \\ \hline
4                         & -0.29              & 0                 & 66                               & 66                              & 66                              & 66                               & 66                               & 82                               \\ \hline
5                         & 0                  & -0.29             & 75                               & 68                              & 165                             & 70                               & 63                               & 114                              \\ \hline
6                         & 0.29               & 0.29              & 86                               & 76                              & 150                             & 82                               & 82                               & 67                               \\ \hline
7                         & -0.29              & 0.29              & 86                               & 78                              & 112                             & 88                               & 92                               & 78                               \\ \hline
8                         & -0.29              & -0.29             & 73                               & 71                              & 154                             & 83                               & 76                               & 140                              \\ \hline
9                         & 0.29               & -0.29             & 78                               & 78                              & 210                             & 85                               & 74                               & 127                              \\ \hline
10                        & 0.15               & 0.15              & 70                               & 68                              & 68                              & 66                               & 68                               & 64                               \\ \hline
11                        & -0.15              & 0.15              & 74                               & 74                              & 68                              & 68                               & 70                               & 65                               \\ \hline
12                        & 0.15               & -0.15             & 62                               & 64                              & 90                              & 65                               & 63                               & 82                               \\ \hline
\multicolumn{3}{|l|}{Field Average}                                & 73                               & 71                              & 109                             & 74                               & 72                               & 85                               \\ \hline
\end{tabular}
\caption{For an input 1.5" (median seeing) beam, radius containing 99.95\% of the ensquared energy of the PSF  for the E and O beams in nominal design and the best and worst case Monte Carlo simulation realizations.}
\label{radii_optical_performance}
\end{table}

\begin{table}[ht!]
\centering
\begin{tabular}{|c|c|c|c|c|c|}
\hline
\multirow{2}{*}{\begin{tabular}[c]{@{}l@{}}Temperature\\ (in Celsius)\end{tabular}} & \multirow{2}{*}{\begin{tabular}[c]{@{}l@{}}Pressure\\ (in Atm)\end{tabular}} & \multicolumn{2}{c|}{O Beams}                                                                                                   & \multicolumn{2}{c|}{E Beams}                                                                                                   \\ \cline{3-6} 
                                                                                    &                                                                              & \multicolumn{2}{c|}{RMS Spot Radius (${\mu}m$)}                                                                                       & \multicolumn{2}{c|}{RMS Spot Radius (${\mu}m$)}                                                                                       \\ \hline
                                                                                    &                                                                              & \begin{tabular}[c]{@{}l@{}}Before \\ Compensation\end{tabular} & \begin{tabular}[c]{@{}l@{}}After \\ Compensation\end{tabular} & \begin{tabular}[c]{@{}l@{}}Before \\ Compensation\end{tabular} & \begin{tabular}[c]{@{}l@{}}After\\  Compensation\end{tabular} \\ \hline
25                                                                                  & 1                                                                            & 11.63                                                          & -                                                             & 11.79                                                          & -                                                             \\ \hline
-10                                                                                 & 0.78                                                                         & 17.1                                                           & 12.5                                                          & 16.7                                                           & 12.08                                                         \\ \hline
0                                                                                   & 0.78                                                                         & 17.3                                                           & 12.19                                                         & 17.05                                                          & 11.94                                                         \\ \hline
10                                                                                  & 0.78                                                                         & 17.55                                                          & 11.92                                                         & 17.36                                                          & 11.84                                                         \\ \hline
20                                                                                  & 0.78                                                                         & 17.7                                                           & 11.76                                                         & 17.65                                                          & 11.78                                                         \\ \hline
30                                                                                  & 0.78                                                                         & 17.9                                                           & 11.7                                                          & 17.9                                                           & 11.77                                                         \\ \hline
\end{tabular}
\caption{Thermal tolerancing results for WALOP-South optical model.}
\label{thermal_tolerancing}
\end{table}


\subsection{Thermal Tolerancing}

While the WALOP-South instrument will be assembled and characterized in the lab at IUCAA at around 1 Atm pressure and $25^{\circ}$~C temperature for which the design has been made, the atmospheric pressure at the Sutherland Observatory is 0.78 Atm due to its higher altitude and the temperature may change from $-10^{\circ}$~C to $30^{\circ}$~C during observations. Changes in pressure and temperature lead to change in thickness, curvature and refractive index of the optics and separation between optical elements due to expansion/contraction of metallic components used as optics holders and spacers. The estimated change in the RMS spot radius for different temperature conditions is shown in the \textit{Before Compensation} column of Table~\ref{thermal_tolerancing}. This image degradation effect can be compensated by adjusting the separation between the  detector and the last lens in each camera. Using Zemax,  thermal tolerancing was carried out on the optical design. As can be seen from the achieved RMS spot radius after compensation in Table~\ref{thermal_tolerancing}, we can get to within 1.1 times the nominal RMS spot radius using this compensator system for all the temperature conditions at the Sutherland Observatory.

\subsection{Stray and Ghost Light Analysis}
\par
\begin{figure}
    \centering
    \frame{\includegraphics[scale = 0.4]{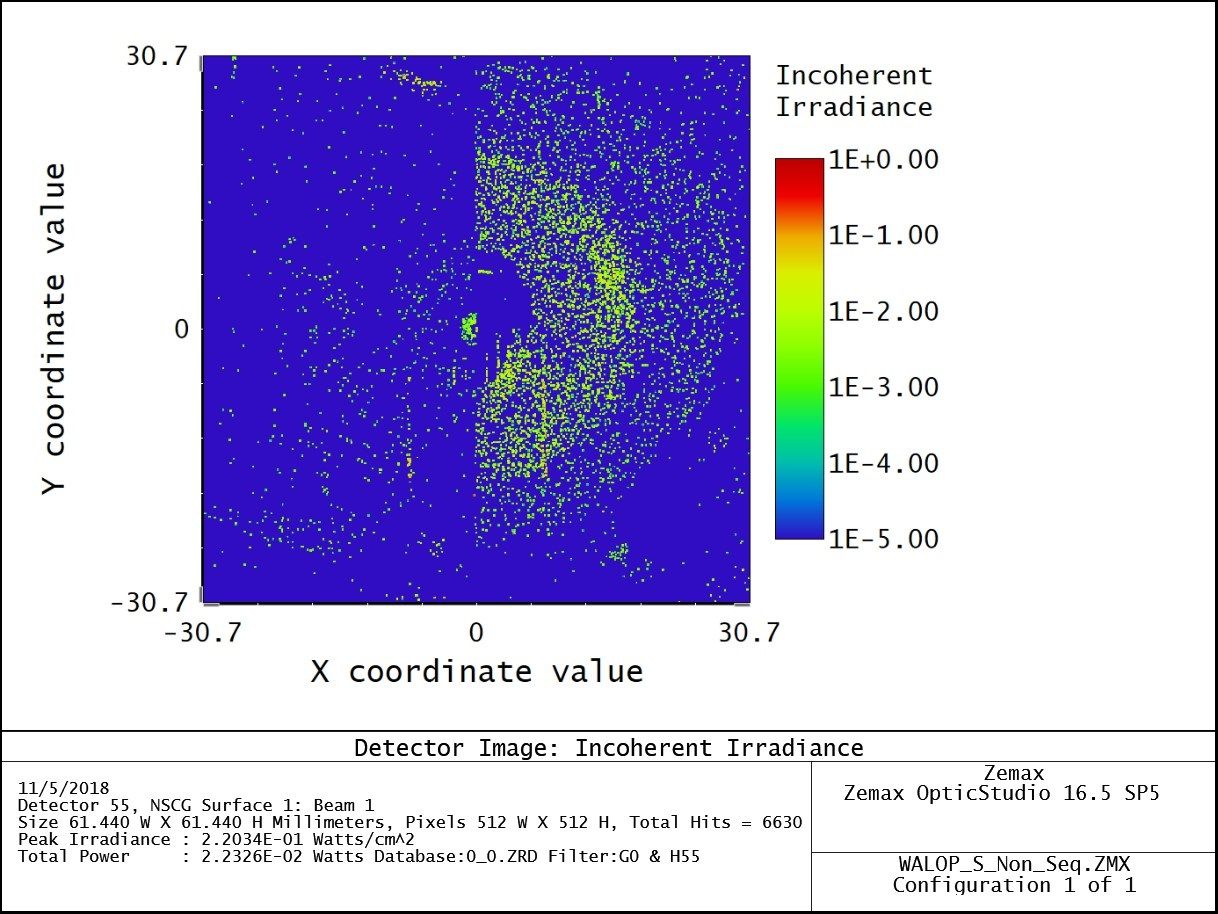}}
    \caption{Ghost image formed for an object at the center of the FOV at one of the four detectors. For a total input power of 100 Watts at the telescope, $2\times10^{-2}$ Watts of power reaches the detector as ghost light.}
    \label{central_ghost}
\end{figure}

 Stray and ghost light analysis for the entire WALOP-South optical system including the telescope mirrors and baffles was carried out using the non-sequential mode of Zemax. As part of the study, we created point sources inside and outside the FOV and light rays were traced through the system. Using Zemax's Path Analysis feature, all paths of stray and ghost light from each source reaching the detectors were identified and controlled in the optical as well as optomechanical
model of the instrument so that for any source the ghost/stray light intensity at the detector is less than $5\times10^{-4}$ of it's total intensity incident on top of the telescope tube. For most sources barring the extremely bright ones ($<10~mag$), its stray and ghost light will be undetectable on WALOP-South detectors as it will be fainter than the background sky. The major source of ghost light from inside FOV object is due to reflection from the lens surfaces at the beginning of the cameras. To control the ghost light as well as improve instrument's throughput, we used high efficiency AR coatings ($R<0.5\%$) for all optical surfaces. Fig~\ref{central_ghost} shows the ghost image pattern from a source at the center of the FOV. For a total input power of 100 Watts incident on top of the telescope tube, $2\times10^{-2}$ Watts of power reaches the detector as ghost image. 
 \par
\begin{figure}
    \centering
    \frame{\includegraphics[scale = 0.4]{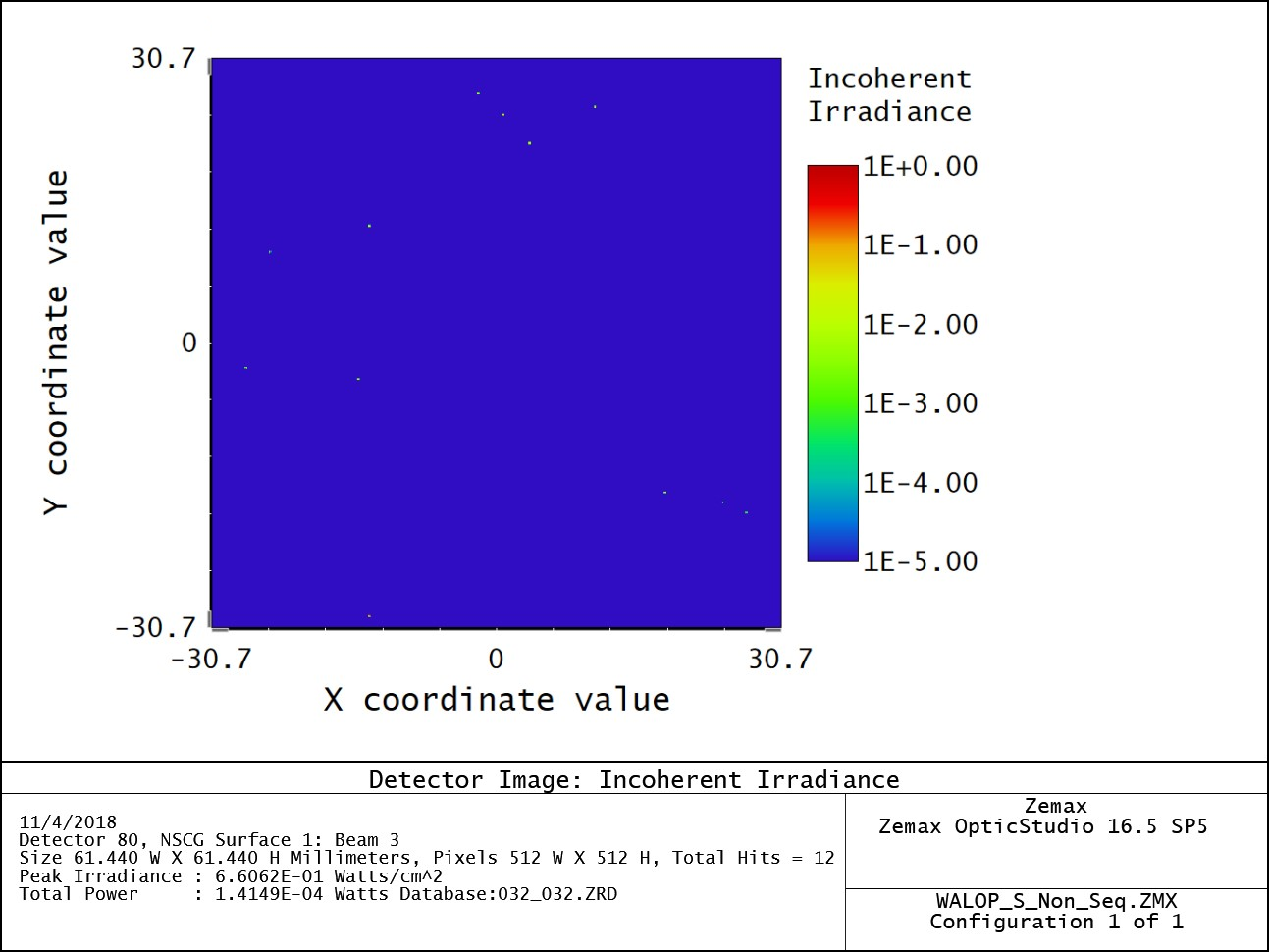}}
    \caption{Stray light irradiation pattern for an object in the vicinity of the FOV (source position: x, y = 19.2, 19.2 arcminute) at one of the four detectors. For a total input power of 100 Watts at the telescope, $1.4\times10^{-4}$ Watts of power reaches the detector.}
    \label{stray1}
\end{figure}

 Fig~\ref{stray1} shows the stray light irradiation for an object in the vicinity of the FOV (source position: x, y = 19.2, 19.2 arcminute) at one of the four detectors. For a total input power of 100 Watts at the telescope, $1.4\times10^{-4}$ Watts of power reaches the detector. 
 
\subsection{Polarimetric Modelling and Calibration}\label{calibration}
To achieve high accuracy linear polarimetry of $0.1~\%$ across the FOV, we need to accurately obtain the value of the real linear Stokes parameters of a star from those measured by the instrument. All optical elements can introduce instrumental polarization and cross-talk in the following ways: (a) due to oblique angles of incidence,  which leads to preferential transmission of one orthogonal polarization over the other , (b) due to stress birefringence in the optics, as a result of thermal and mechanical stresses on the optics. Over and above these, we expect the main source of instrumental cross-talk to arise from the non half-wave retardance from the HWPs in WPA, as described in Sec~\ref{WPA Architecture}.
\par So the measured value of the linear Stokes parameters by the instrument will depend on the input polarization of the source (cross-talk) and the instrumental zero point (instrumental polarization). To correct for these effects and obtain high accuracy polarization values, we have developed a detailed model for on-sky calibration of the instrument. For this purpose, we have placed a movable linear polarizer at the beginning of the instrument (after the telescope optics and before the first collimator lens), which will be removed from the optical path during the main observations. This polarizer will be used to provide as input linearly polarized light with different EVPAs to the instrument, which will be used to create a mapping function between the instrument's measured and real polarization values of stars. Additionally, the model uses standard polarized and unpolarized stars for building as well as testing the accuracy of the model. We have tested and validated this scheme by using the Zemax optical model of the instrument and have achieved better than 0.1~\% accuracy over the entire FOV. We will test the calibration model on sky during the commissioning of WALOP-South instrument and details of the model and results obtained will be published as a separate paper. 

\color{black}
\section{Conclusions}\label{conclusion}

\par We have described the complete optical model of the WALOP-South instrument which meets all the design goals to successfully carry out the PASIPHAE Survey. Scheduled for commissioning in 2021, it will be a unique wide field polarimeter. In one shot, it allows determination of I , $q$ and $u$ with four channels simultaneously imaged on separate detector/camera operating over a broadband filter wavelength range. We have elaborated the key challenges in creating the design, namely the dispersion introduced by large split angle Wollaston Prisms and aberrations due to the very wide field. With WALOP-South optical design, we expect to obtain within 1.6 times the seeing limited PSF for the entire FOV after correcting for these effects. 

\par While we have developed the optical model to work for the SDSS-r filter and narrowband filters within the wavelengths of $0.5~{\mu}m - 0.7~{\mu}m$, the prescription presented can be implemented with minor modifications to design polarimeters to work over other broadband filters. It can also be used to design polarimeters with larger FOV by using WPs with larger aperture than those used in WALOP-South, most likely by employing a mosaic of smaller WPs, without increasing the split angle and associated spectral aberrations from the WPs. Also, of interest is the possibility of carrying out low-resolution imaging spectropolarimetry in WALOP-South like polarimeters by taking advantage of the dispersion created by WPs by not correcting for the dispersion.

\appendix
\section{Baffles Design}\label{baffles}
\par
\begin{figure}
\centering
\includegraphics[scale = 0.5]{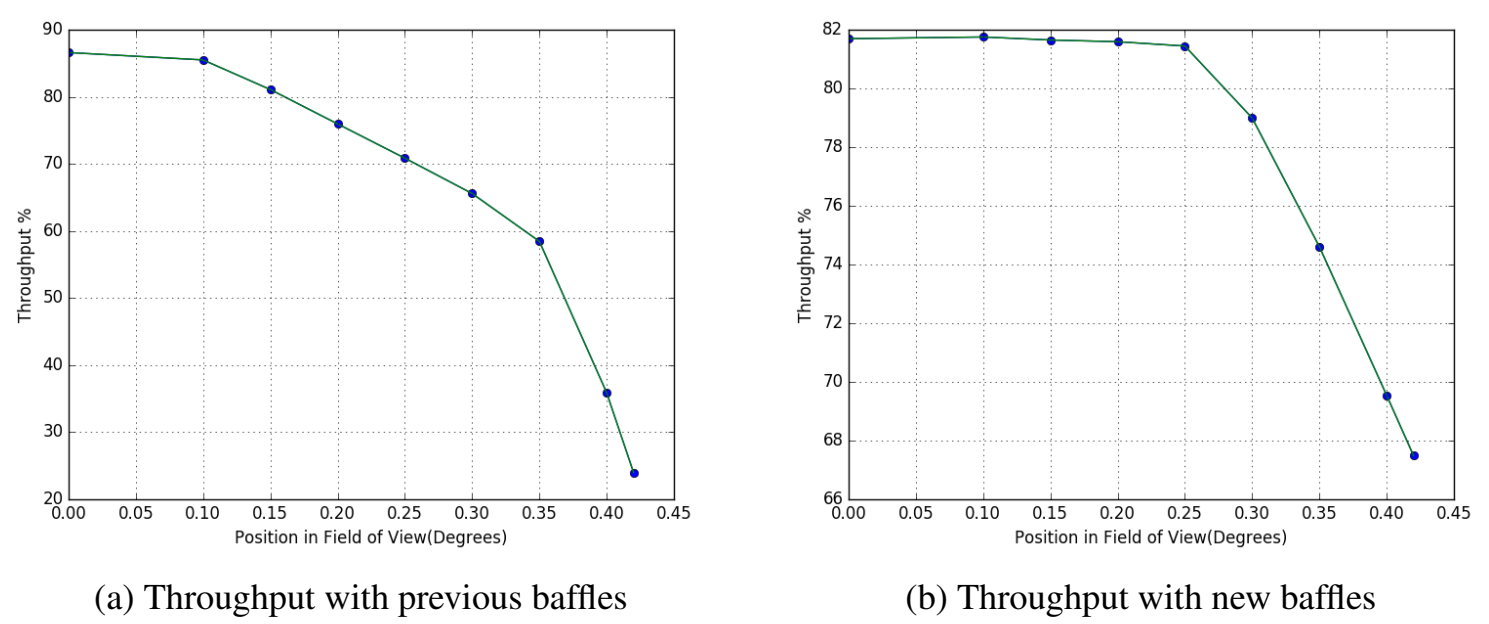}
\caption{Comparison of throughput at the telescope focal plane between the previous baffles and the new baffles for the telescope designed to accommodate WALOP-South's FOV.}
\label{baffle_throughput}
\end{figure}

 Telescope baffles are used to prevent direct stray light from sources outside the FOV of the instrument to reach the telescope focal plane and propagate into the instrument. An efficient baffle system should achieve this objective with minimal obstruction to light from sources inside the FOV. The existing baffles at the 1~m SAAO telescope were optimized for a narrow FOV of 12 arcminutes in diameter. The throughput with these baffles at the telescope focal plane for the radial extent of WALOP-South FOV (the extreme field point is at a distance of $0.41^{\circ}$ from the center) is shown in Fig~\ref{baffle_throughput}~a. While this baffle allows high throughput for objects up to a radius of 6 arcminutes, there is a steep drop in throughput for farther objects. In order to get uniform throughput for the entire WALOP-South FOV, we designed a new baffle system for the telescope's primary and secondary mirrors using the method described by Kumar et al.\cite{SenthilKumar2} Fig~\ref{baffle_throughput}~b shows the throughput at the telescope focal plane with the new baffle system. Table~\ref{30_arcmin_baffle} captures the dimensions of the new baffles. The new baffle system has been optimised for  a FOV of 30~arcminutes in diameter, and there is minor drop in throughput for farther field points. The average throughput for the complete field is 80.3\% in comparison to 72.8\% from the previous baffles. 

\begin{table}
\centering
 \begin{tabular} {|c|c|c|}
 \hline
 \textbf{Baffle} & \textbf{Length}~(mm) & \textbf{Outside Diameter}~(mm) \\
 \hline
 Primary & 1673.6 & 236.2 \\
 \hline
 Secondary &  610.3 & 424.2 \\
 \hline
 \end{tabular}
\caption{Dimensions of the new telescope baffles designed for the 1~m SAAO telescope.}
\label{30_arcmin_baffle}
\end{table}

\section{Guider Camera Design}\label{guider}
\par
\begin{figure}
\centering
\includegraphics[scale=0.5]{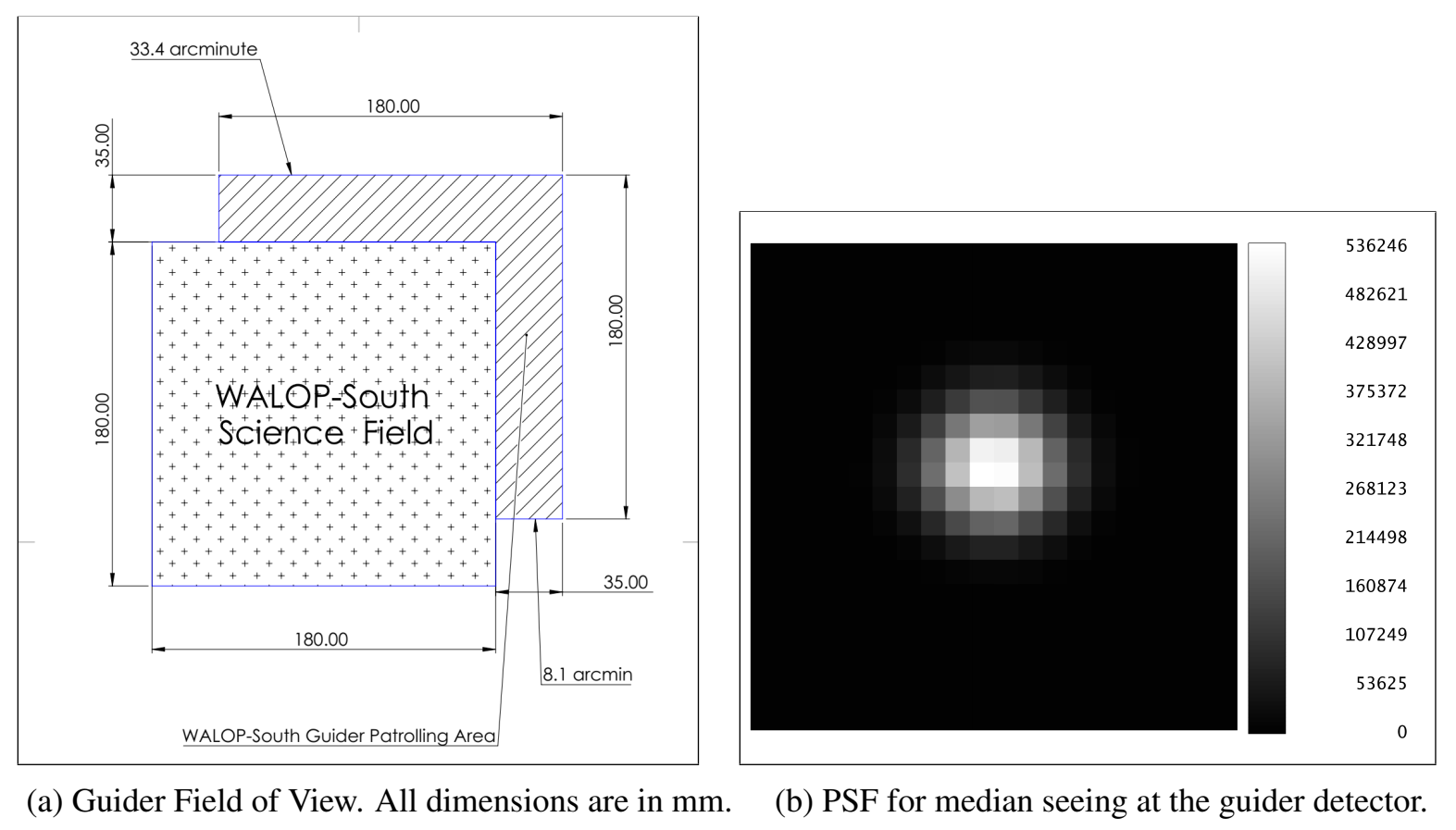}

\caption{(a) Schematic of the available FOV for WALOP-South guider camera and (b) PSF at the guider camera detector across the FOV.}
\label{guider_camera}
\end{figure}

 As WALOP-South will be mainly observing the southern Galactic polar regions where there are less number of stars available per square degree, there is sparsity of available guide stars. Additionally, the guider camera field must lie outside of the main science field as any guider optics and associated mechanical structure may lead to reflection of light leading to stray light on the detector. As can be seen in Fig~\ref{baffle_throughput}, the throughput from the telescope is lower outside the science field. The auto-guider camera for the instrument needs to have a large enough FOV to be able to find guide stars for all fields to be covered in PASIPHAE survey. We estimate a minimum required FOV area of 200 square arcminutes for the guider camera. The guider camera designed for WALOP-South is a modified version of the guider camera on the \href{https://www.saao.ac.za/explore/our-telescopes/lesedi/}{1~m Lesedi Telescope} at SAAO's Sutherland Observatory. Fig~\ref{guider_camera}~a shows the FOV for the guider camera which is located near the telescope focal plane. This region spanning 540 square arcminutes will be patrolled using two linear stages on which the camera optics, consisting of a pick-off mirror and a lens doublet, and a Lodestar X2 CCD detector will be mounted. The pick-off mirror will be moved to the position of the light beam path coming from the target guide star and folded sideways to be imaged on the detector after correction of aberrations by the lens doublet. Fig~\ref{guider_camera}~b shows the PSF obtained (FWHM = 2.2") for median seeing FWHM of 1.5".

\appendix

\acknowledgments 
\par The PASIPHAE program is supported by grants from the European Research Council (ERC) under grant agreement No 771282 and No 772253, from the National Science Foundation, under grant number AST-1611547 and the National Research Foundation of South Africa under the National Equipment Programme. This project is also funded by an infrastructure
development grant from the Stavros Niarchos Foundation and from the Infosys Foundation.

\par We are thankful to Vinod Vats at Karl Lambrecht Corp. for his inputs and suggestions on various aspects of the polarizer assembly design and fabrication and Prof. Kenneth Nordsieck for sharing his experience and ideas on thermal properties of calcite Wollaston Prisms.

\par We thank the anonymous reviewers of the paper whose comments and suggestions helped improve the paper.


\bibliography{WALOP_S_Optical_Design_paper}   
\bibliographystyle{spiejour}   


\vspace{2ex}\noindent\textbf{Siddharth Maharana} is an astrophysics PhD student at the Inter-University Centre for Astronomy and Astrophysics, Pune, India. He received his Bachelor in Mechanical Engineering from Central University, Bilaspur, India in 2015. He is currently working on the design and development of the WALOP instruments for PASIPHAE survey. His areas of interest are polarimetric instrumentation and data analysis.

\vspace{2ex}\noindent\textbf{John A. Kypriotakis} is a Ph.D. student at the University of Crete, Greece, Department of Physics and at the Institute for Astrophysics of the Foundation for Research and Technology Hellas, Greece. He received his B.Sc. in Physics from University of Crete, Greece in 2017. He is currently working on the design and development of the WALOP instruments for the PASIPHAE survey. His areas of interest are Instrumentation (incl. Software), Data Analysis and Machine Learning.

\vspace{2ex}\noindent\textbf{A. N. Ramaprakash} Not Available

\vspace{2ex}\noindent\textbf{Chaitanya Rajarshi} Not Available

\vspace{2ex}\noindent\textbf{Ramya M. Anche} Not Available

\vspace{2ex}\noindent\textbf{Shrish} Not Available

\vspace{2ex}\noindent\textbf{Dmitry Blinov} Not Available

\vspace{2ex}\noindent\textbf{Hans Kristian Eriksen} Not Available

\vspace{2ex}\noindent\textbf{Tuhin Ghosh} Not Available

\vspace{2ex}\noindent\textbf{Eirik Gjerløw} Not Available

\vspace{2ex}\noindent\textbf{Nikolaos Mandarakas} Not Available

\vspace{2ex}\noindent\textbf{Georgia V. Panopoulou} Not Available

\vspace{2ex}\noindent\textbf{Vasiliki Pavlidou} Not Available

\vspace{2ex}\noindent\textbf{Timothy J. Pearson} Not Available

\vspace{2ex}\noindent\textbf{Vincent Pelgrims} Not Available

\vspace{2ex}\noindent\textbf{Stephen B. Potter} Not Available

\vspace{2ex}\noindent\textbf{Anthony C. S. Readhead} Not Available

\vspace{2ex}\noindent\textbf{Raphael Skalidis} Not Available

\vspace{2ex}\noindent\textbf{Konstantinos Tassis} Not Available

\vspace{2ex}\noindent\textbf{Ingunn K. Wehus} Not Available
\vspace{1ex}

\listoffigures
\listoftables

\end{document}